\documentclass[10pt,journal,compsoc]{IEEEtran}
\ifCLASSOPTIONcompsoc
  \usepackage[nocompress]{cite}
\else
  \usepackage{cite}
\fi

\usepackage{xcolor}
\usepackage{array}
\usepackage{subfigure}
\usepackage{graphicx}
\usepackage{grffile}
\usepackage{amssymb}
\usepackage{amsmath}
\usepackage{amsfonts}
\usepackage{amsthm}
\usepackage{multirow}
\usepackage{color}
\usepackage{url}
\usepackage{cite}
\usepackage[lined,linesnumbered,ruled]{algorithm2e}
\usepackage{todonotes}
\setuptodonotes{inline}

\newtheorem{lemma}{Lemma}

\newtheorem{theorem}{Theorem}

\newtheorem{definition}{Definition}

\begin{document}

\title{PDSR: A Privacy-Preserving Diversified Service Recommendation Method on Distributed Data}

\author{Lina~Wang,
        Huan~Yang,
        Yiran~Shen,
        Chao~Liu,
        Lianyong~Qi,
        Xiuzhen~Cheng,
        and~Feng~Li 
\IEEEcompsocitemizethanks{
\IEEEcompsocthanksitem L. Wang, X. Cheng, and F. Li are with School of Computer Science and Technology, Shandong University, Qingdao, 266237, China. \protect\\
E-mail: linawang425@mail.sdu.edu.cn, \{xzcheng, fli\}@sdu.edu.cn. 
\IEEEcompsocthanksitem H. Yang is with College of Computer Science and Technology, Qingdao University, Qingdao, 266071, China. \protect\\
Email: cathy\_huanyang@hotmail.com.
\IEEEcompsocthanksitem Y. Shen is with School of Software, Shandong University, Jinan, 250101, China. \protect\\
E-mail: yiran.shen@sdu.edu.cn. 
\IEEEcompsocthanksitem C. Liu with Department of Computer Science and Technology, Ocean University of China, Qingdao, 266100, China. \protect\\
E-mail: liuchao@ouc.edu.cn
\IEEEcompsocthanksitem L. Qi is with College of Computer Science and Technology, China University of Petroleum (East China), Qingdao, 266580, China.  \protect\\ 
E-mail: 20220115@upc.edu.cn.}
}
%



\IEEEtitleabstractindextext{%
\begin{abstract}
 The last decade has witnessed a tremendous growth of service computing, while efficient service recommendation methods are desired to recommend high-quality services to users. It is well known that collaborative filtering is one of the most popular methods for service recommendation based on QoS, and many existing proposals focus on improving recommendation accuracy, i.e., recommending high-quality redundant services. Nevertheless, users may have different requirements on QoS, and hence diversified recommendation has been attracting increasing attention in recent years to fulfill users' diverse demands and to explore potential services. Unfortunately, the recommendation performances relies on a large volume of data (e.g., QoS data), whereas the data may be distributed across multiple platforms. Therefore, to enable data sharing across the different platforms for diversified service recommendation, we propose a \textit{ Privacy-preserving Diversified Service Recommendation} (PDSR) method. Specifically, we innovate in leveraging the Locality-Sensitive Hashing (LSH) mechanism such that privacy-preserved data sharing across different platforms is enabled to construct a service similarity graph. Based on the similarity graph, we propose a novel accuracy-diversity metric and design a $2$-approximation algorithm to select $K$ services to recommend by maximizing the accuracy-diversity measure. Extensive experiments on real datasets are conducted to verify the efficacy of our PDSR method.
 
\end{abstract}

}

\maketitle

\IEEEdisplaynontitleabstractindextext

%
\IEEEpeerreviewmaketitle

\section{Introduction}\label{sec:intro}
  Recent years have witnessed a considerable development of services, thanks to the emerging computing paradigms and architectures, e.g., \textit{Software Oriented Architecture} (SOA), Internet of Services (IoS) and Cloud Computing. As a growing variety of services are provided on the Web, it is of considerable importance to recommend appropriate services for a wide range of users~\cite{GhafouriHH-TSC21}.

  It is well known that \textit{Collaborative Filtering} (CF) has been widely used to build service recommender systems based on \textit{Quality of Service} (QoS)~\cite{LindenSJ-IntComp03,Hadad-TSC10,MaWHHSY-TSC15}. The basic idea is to exploit the similarity among services to predict QoS values for particular users, while characterizing the service similarity relies on mining a big volume of historical QoS data. 
  Traditional recommendation methods usually aim at improving recommendation accuracy such that they recommend only similar services with the highest predicted QoS to target users, whereas the users may have different requirements on the QoS~\cite{ZhangWHCDY-ICWS19,KangLCX-SCC20}. Furthermore, the recommendation highly relies on the QoS prediction; nevertheless, the prediction usually has inevitable bias. Therefore, recommendation diversity is extensively studied in many recent proposals~\cite{ZhengGCJL-WWW21,KangTLLC-TSC16,YeHLZZGWL-RecSys21,LiuXWMZZT-AAAI20,WuKLHQW-WCMC20}. On one hand, it is able to fulfill the users' different QoS demands; on the other hand, it helps to explore potential services which is underestimated by the QoS prediction.

  Unfortunately, diversifying service recommendation usually entails a large volume of data (e.g., QoS data), whereas the data may be distributed across different platforms~\cite{QiZDN-JSAC17}. Despite the benefits of enabling data sharing across the different platforms, the platforms owing the data are usually reluctant to share their data with each other, since private information could be inferred from the data by untrusted third-party platforms~\cite{ZhuHZL-ICWS15,QiXDYQZ-ICWS17}. Therefore, to enabling data sharing across different platforms, privacy preservation is one of the main concerns. Many efforts have been made to integrate privacy preservation techniques with CF. For example, \cite{Canny-SP02,JumonjiSSK-ICDE22} apply homomorphic encryption techniques, and differential privacy mechanisms are adopted in \cite{ZhuLRZX-ASONAM13,GaoHLJL-SIGIR20} to obfuscate the original data before sharing them with others. Nevertheless, the cryptography-based method entails high computation overhead, while the differential privacy mechanism usually impairs data utility. Another choice is to utilize hashing mechanisms. For example, \textit{Locality-Sensitive Hashing} (LSH), have been utilized in \cite{ChowPW-ICDMW12,AhgasaryanBKKN-TrustCom13,QiXDYQZ-ICWS17} to ensure privacy preservation based on distributed data sources. It is usually applied to efficiently exploit the similarity among the services or users through light-weight computations, while the privacy preservation can be ensured thanks to its irreversibility. Unfortunately, how to effectively diversify the similarity relationship obtained by LSH is still an open problem.

  In this paper, we propose a \textit{Privacy-preserving Diversified Service Recommendation} (PDSR) method. The proposed PDSR method not only leverages a trade-off between recommendation accuracy and diversity, but also ensures privacy preservation for data sharing across different platforms. In particular, LSH mechanism is adopted to construct a graph to characterize the similarity among the services based on distributed data from different sources. Due to the irreversibility and the computational efficiency of the hash functions, the construction of the similarity graph entails light-weight computational overhead and enables privacy preservation across the different data sources. We innovate in designing a new diversity metric for the similarity graph, based on which, we formulate our $K$-\textit{Diversified Service Recommendation} ($K$-DSR) problem where $K$ services are selected to maximize the accuracy-diversity evaluation of the recommendation. We prove the NP-hardness of the problem, and design a $2$-approximation algorithm thanks to its monotone submodularity.

  In summary, we make the following main contributions in our PDSR method as follows:
  \begin{itemize}
    \item We leverage the notion of LSH to efficiently construct a weighted service similarity graph across different data sources with their privacy preserved.
    \item We innovate in designing a new diversity metric for the similarity graph, based on which, a $2$-approximation algorithm is proposed for diversified recommendation.
    \item We finally perform extensive experiments on real data to verify the efficacy of our proposed method in leveraging the accuracy-diversity trade-off. 
  \end{itemize}

  The remaining of our paper is organized as follows. We first survey related literature and present an example to motivate the design of our service recommendation method in Sec.~\ref{sec:relwk}. We then give some preliminaries in  Sec.~\ref{sec:pre}. Moreover, we report the details of our proposed PDSR method in Sec.~\ref{sec:metd}. Experiment results are then given in Sec.~\ref{sec:experiment}. We finally conclude this paper in Sec.~\ref{sec:conclusion}.

\section{Related work and Motivation} \label{sec:relwk}
  \subsection{Diversified Service Recommendation} \label{ssec:relwk-divrec}

    Some of state-of-the-art proposals mainly focus on developing innovative methods to enhance the result accuracy \cite{ZhongFHTZ-ICWS14,MaWHHSY-TSC15,ZhangWHCDZY-TSC19,LiWPPYC-KBS19,CuiXXCCZC-TSC20}. For example, in \cite{ZhangWHCDZY-TSC19}, a covering-based service recommendation method is proposed through neighbor-aware matrix factorization. \cite{LiWPPYC-KBS19} designs a deep learning-aided collaborative filtering framework, where neural networks is adopted to learn binary representations for both users and items. In \cite{CuiXXCCZC-TSC20}, time correlation coefficient and $K$-means clustering with cuckoo search are employed to improve the performance of collaborative filtering by taking into account users' time-varying interests. 

    As mentioned in Sec.~\ref{sec:intro}, an accurate recommendation is not necessarily a satisfactory one. Therefore, diversified recommendation have been attracting increasing attention in recent years. In \cite{KangTLLC-TSC16}, functional relevance, QoS utility, and diversity features of Web services are incorporated for recommending well diversified services to users. In \cite{ChenZZ-NIPS18}, an acceleration algorithm is proposed to speed up the \textit{Maximum A Posteriori} (MAP) inference for determinantal point process; the algorithm can be used for recommendation diversification. The accelerated MAP inference algorithm is then adopted in \cite{LiuXWMZZT-AAAI20}. In particular, \cite{LiuXWMZZT-AAAI20}, the fast MAP inference algorithm is utilized in a combinatorial bandit for diversified recommendation. \cite{ZhangWHCDY-ICWS19} exploits the quality correlations among different dimensions of users' quality preferences, to diversify recommendation results. In \cite{LathaN-Physica19}, user exposure diversity and item concordance are exploited to endow the collaborative filtering with recommendation diversity. In \cite{WuKLHQW-WCMC20}, multiple recommendation lists are selected first, from which the most diversified one can be found according to a list diversity measure. In \cite{ZhengGCJL-WWW21}, rebalanced neighbor discovering, category boosted negative sampling and adversarial learning are performed on top of graph convolutional networks, so as to realize diversity in recommendation. In \cite{YeHLZZGWL-RecSys21}, an end-to-end dynamic diversified graph framework is proposed to construct the user-item graph dynamically based on the users and items embeddings, and a quantile progressive candidate selection operator is designed to efficiently select diverse items to recommend.

  \subsection{Privacy-Preserved Service Recommendation} \label{ssec:privacy}

    The above studies rely on a huge volume of data, by exploiting which recommendation results are diversified. However, data are usually stored on different platforms, while data sharing across the different sources may considerably increases the risk of privacy leakage. Hence, one of another main concerns in service recommendation is the privacy issue. In \cite{DouZLC-TPDS15}, only a small fraction of data are shared publicly to reduce the disclosure of sensitive data. By this way, although privacy is preserved, but it is with a significant sacrifice in recommendation performance. Different from \cite{DouZLC-TPDS15}, \cite{ZhuHZL-ICWS15} applies data obfuscation for the purpose of privacy preservation. Specifically, data are obfuscated through randomization techniques before being shared for recommendation. Likewise, the notion of differential privacy is leveraged in \cite{LiPJ-BigData17}; the data privacy is preserved through insert ``noise'' into the data before sharing them. However, in the above two methods, data are obfuscated or noised such that their utility is impaired. Additionally, when the data are stored in different platforms, sharing the obfuscated (or noised) data across different platforms still results in considerable communication overhead.

    Another choice for privacy preservation is to utilize \textit{Locality-Sensitive Hashing} (LSH). In \cite{QiXDYQZ-ICWS17}, the similarity retention is LSH is employed such that different platforms can share the similarity among the services (which can be indicated by the hash values) with each other without disclosing the original data. Furthermore, the LSH-based method is developed by \cite{ZhangYZLDWQ-COMCOM20} by exploring mutiple dimensions of QoS data. Although these proposals have made considerable effort in combining LSH and CL for privacy-preserved recommendation, they concentrate in pursuing high recommendation accuracy and do not take into account recommendation diversity. Although the LSH mechanism is used to calculate both similarity and diversity in \cite{WangZWYKQ-KBS20}, this proposal does not exploit the application of the LSH mechanism in privacy-preservation across different data sources. Moreover, \cite{WangZWYKQ-KBS20} calculates the diversity directly according to the LSH values, while we exploit the service similarity graph to define the recommendation diversity and gain better recommendation results, as will be shown in Sec.~\ref{sec:metd}.

  \subsection{Motivation} \label{ssec:motivation}
\begin{figure}[htb]
    \begin{center}
      \includegraphics[width=0.89\columnwidth]{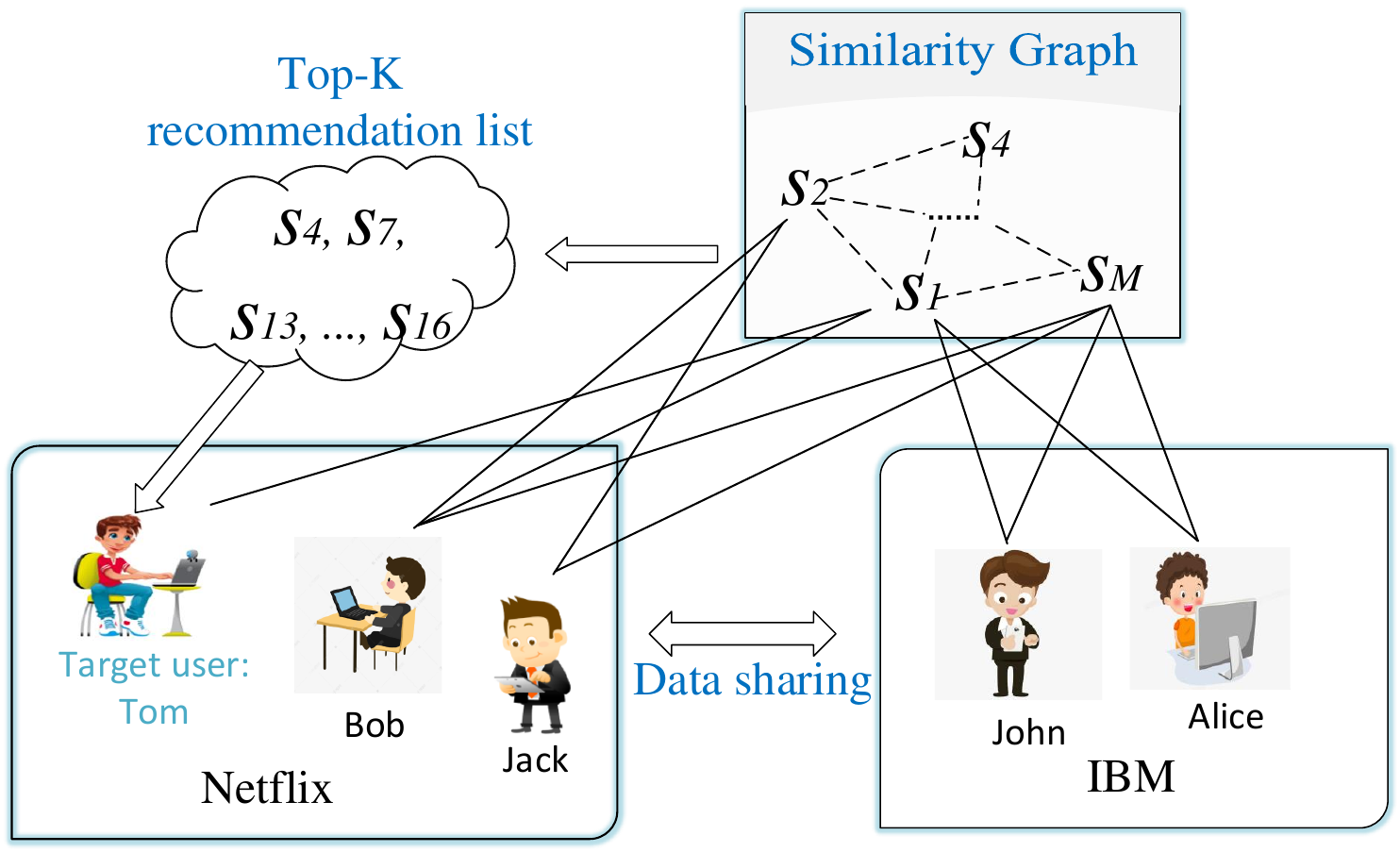}
    \caption{Service recommendation across different data sources.}
    \label{fig:mot}
    \end{center}
    \end{figure}

    We hereby give an example to explain the application scenario of our service recommendation method. As shown in Fig.~\ref{fig:mot}, we assume that there are two platforms (or data sources), i.e., Netflix and IBM. We also suppose that there are three users, i.e., Tom, Bob and Jack, on Netflix platform and another two users John and Alice on IBM platform. Let $s_1, s_2, \cdots, s_M$ denote $M$ different services invoked by these users, respectively. Note that, a service may be invoked by multiple users on different platforms. For example, $s_1$ is invoked by Tom and Bob on Netflix platform and by John and Alice on IBM platform, and $s_M$ is invoked by Bob and Jack on Netflix platform and by John and Alice on IBM platform. The corresponding QoS data are then stored on different platforms. We suppose Netflix intends to recommend new services to Tom. Through the CF method, Netflix first computes the similarity (e.g., by constructing a similarity graph) among the services according to its local QoS data and then seeks for the services that are similar to the historical records of Tom. Specifically, for any new service for Tom, its QoS value can be predicted according to the ones of the other similar services. Usually, $K$ services with the highest predicted QoS values are finally selected and recommended to Tom.

    As mentioned above, there are two challenging issues in the above recommendation process. On one hand, similar services are recommended to Tom in order to ensure recommendation accuracy, while the redundancy of the services may impair the experience of the recommendation for Tom. On the other hand, although data sharing between Netflix and IBM will be helpful for data-driven recommendation diversification, enabling the data sharing across different platforms efficiently is highly non-trivial especially due to privacy concerns. Therefore, to tackle the above challenges, we first leverage the notion of LSH to enable efficient data sharing between Netflix and IBM, based on which, we can characterize the similarity among the services more comprehensively. Furthermore, we propose a new diversity metric such that Netflix can mine diversified recommendation for Tom through the cumulative data more effectively. 

\section{Preliminaries} \label{sec:pre}
  Before diving into the design and analysis of our proposed algorithm, we first introduce some preliminaries in this section, which will be very helpful later.
  \subsection{Locality-Sensitive Hashing} \label{ssec:lsh}
    LSH is an efficient approximate nearest neighbor search method \cite{GionisIM-VLDB99}. Given a set of points (e.g., services which can be represented as a high-dimensional vectors or points in our case), the goal of traditional hashing methods is to map the points to a set of values such that the points can be ``spread'' randomly, while LSH aims at mapping the points with the so-called ``locality'' guaranteed. Specifically, by LSH, the points which are ``closed'' to each other in the original data space can be hashed to the same value with a high probability, whereas there is a high probability for the ones ``far away'' from each other to be hashed to different values. In another word, it is more likely for ``similar'' points to have the same hash value.

    Let $\mathsf{Dis}(s, s')$ denote the ``distance'' between the points $s$ and $s'$. The distance between $s$ and $s'$ is smaller when $s$ and $s'$ are more similar to each other. We formally define LSH in the following \textbf{Definition}~\ref{def:lsh}.
    \begin{definition}[Locality Sensitive Hashing] \label{def:lsh}
      Given a set of points $\mathcal S$, let $\mathcal H = \{h \mid \mathcal S \rightarrow \mathbb Z^+\}$ be a family of hash functions which maps the points $\mathcal S$ to a set of hash values $\mathbb Z^+$. $\mathcal H$ is said to be a $(\varepsilon_1, \varepsilon_2, p_1, p_2)$-LSH if
      \begin{equation} \label{eq:lsh}
        \begin{cases}
          \mathsf{Pr}_{\mathcal H}(h(s)=h(s')) \geq p_1,~~\text{if}~\mathsf{Dis}(s, s') \leq \varepsilon_1 \\
          \mathsf{Pr}_{\mathcal H}(h(s)=h(s')) \leq p_2,~~\text{if}~\mathsf{Dis}(s, s') \geq \varepsilon_2 
        \end{cases}
      \end{equation}
      holds for any $s, s' \in \mathcal S$, where $p_1 > p_2$. The size of the resulting hash table is the number of integers in $\mathbb Z^+$, and we also suppose $H$ is the number of functions in $\mathcal H$. Note that the probability is over the random choices of $h \in \mathcal H$.
    \end{definition}

    It is shown by the above definition that, in the LSH mechanism, similar points are mapped to the same value with high probability; therefore, we can figure out the similarity among any points (i.e., the services we intend to recommend) according to their hash values instead of their coordinates in the original data space (i.e., the private information of the services). In another word, for any services which are mapped to the same value through LSH, they are similar to each other with high probability. Furthermore, the hash functions are irreversible such that we can observe the similarity among the points without disclosing their ``private'' coordinates. As will be shown in Sec.~\ref{sec:metd}, we leverage the notion of LSH to construct a graph to characterize the similarity among the services by enabling privacy-preserved data sharing across different platforms.

  \subsection{Expanded Set and Expansion Ratio} \label{ssec:expansion}
    Let $\mathcal G=(\mathcal M, \mathcal E)$ denote a graph, where $\mathcal M$ and $\mathcal E$ denote the set of vertices and the set of edges, respectively. We define expanded set and expansion ratio for any subset $\mathcal M'\subseteq \mathcal M$ as follows.
    \begin{definition}[Expanded set] \label{def:expset}
      Given a graph $\mathcal G=(\mathcal M, \mathcal E)$, for any subset of vertices $\mathcal M' \subseteq \mathcal M$, the expanded set of $\mathcal M'$ can be defined as 
      \begin{equation} \label{eq:expset}
        \mathsf{Exp}(\mathcal M') = \mathcal M' \bigcup \left\{ i \in \mathcal M/ \mathcal M' \mid \exists i'\in\mathcal M', (i,i')\in\mathcal E  \right\}
      \end{equation}
    \end{definition} 
    \begin{definition}[Expansion ratio] \label{def:expratio}
      Let $\lvert \mathsf{Exp}(\mathcal M') \rvert$ denote the number of vertices in the expanded set of $\mathcal M' \subseteq \mathcal M$. We also suppose $M = |\mathcal M|$ denotes the number of vertices in graph $\mathcal G$. The expansion ratio of $\mathcal M'$ can be defined by
      \begin{equation} \label{eq:expratio}
        \alpha(\mathcal M') = \frac{\lvert \mathsf{Exp}(\mathcal M') \rvert}{M}
      \end{equation}
    \end{definition}
    
\begin{figure}[htb!]
    \centering
      \subfigure[Graph $\mathcal G$]{
        \begin{minipage}{0.3\linewidth}
           \centering
           \includegraphics[width=1in]{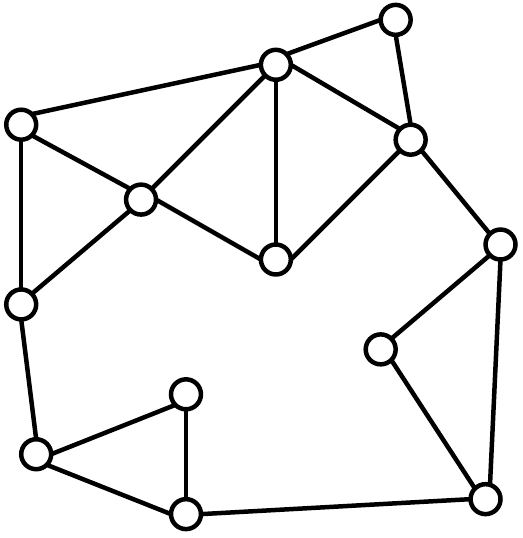}
        \vspace{0.02cm} 
        \end{minipage}}
      \subfigure[$\alpha(\mathcal M'_1)=8/13$]{
        \begin{minipage}{0.3\linewidth}
           \centering
           \includegraphics[width=1in]{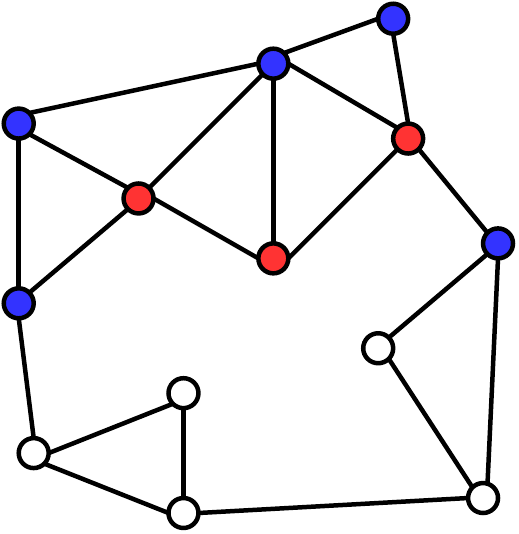}
        \vspace{0.02cm} 
        \end{minipage} }
      \subfigure[$\alpha(\mathcal M'_2)=10/13$]{
        \begin{minipage}{0.3\linewidth}
           \centering
           \includegraphics[width=1in]{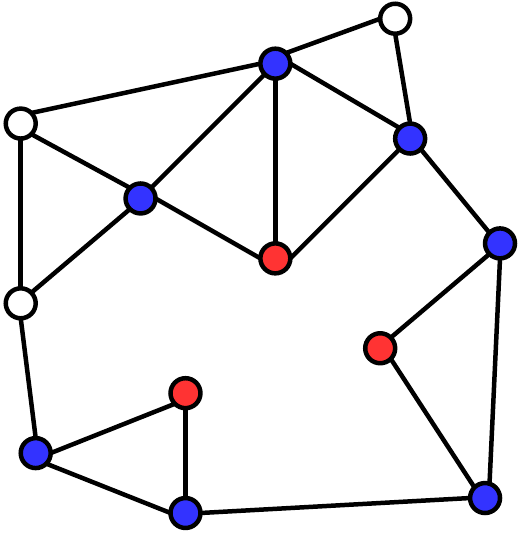}
        \vspace{0.02cm} 
        \end{minipage}}
    \caption{An illustration of expanded set and expansion ratio.}
    \label{fig:expand}
    \end{figure}

    According to the above definition, the expansion radio of a subset $\mathcal M' \subseteq \mathcal M$ actually measures its trend of expanding outwards. If we let the size of $\mathcal M'$, i.e., $M'=|\mathcal M'|$, be fixed, a larger expansion ratio of $\mathcal M'$ implies that the vertices in $\mathcal M'$ have more neighbors outside (i.e., in $\mathcal M / \mathcal M'$), and hence it is less likely for the vertices in $\mathcal M'$ to share the same neighbors in $\mathcal M/\mathcal M'$. For example, as illustrated in Fig.~\ref{fig:expand}, we use red vertices and blue vertices to represent the selected subset and its expanded neighbors, respectively. The subset $\mathcal{M}'_1$ (consisting of the three red vertices) in Fig.~\ref{fig:expand} (b) has an expansion ratio $\alpha(\mathcal{M}'_1)=8/13$ and the three vertices share the same neighbor, while its counterpart $\mathcal{M}'_2$ in Fig.~\ref{fig:expand} (c) has an expansion ratio $\alpha(\mathcal{M}'_2)=10/13$ and none of the three red vertices in $\mathcal{M}'_2$ share the same neighbor, even $\mathcal{M}'_1$ and $\mathcal{M}'_2$ have the same number of outgoing edges.

  \subsection{Submodularity} \label{ssec:mod}
    The concept of submodular set function is very useful in approximation algorithms. In the following, we present a definition of monotone submodular set function, based on which we can design and analyze our graph-based recommendation algorithm. 
    \begin{definition}[Monotone submodular function] \label{def:modufunc}
      Given a finite set $\mathcal M$, let $\mathsf{F}: 2^{\mathcal M} \rightarrow \mathbb R$ be a real valued set function. $\mathsf{F}(\cdot)$ is said to be a monotone submodular set function, if the following conditions meet.
      \begin{itemize}
        \item Monotonicity. For any two subsets $\mathcal M'_1, \mathcal M'_2 \subseteq \mathcal M$ such that $\mathcal M'_1 \subseteq \mathcal M'_2 \subseteq \mathcal M$, we have $\mathsf{F}(\mathcal M'_1) \leq \mathsf{F}(\mathcal M'_2)$.
        \item Submodularity. For any two subsets $\mathcal M'_1, \mathcal M'_2 \subseteq \mathcal M$ such that $\mathcal M'_1 \subseteq \mathcal M'_2 \subseteq \mathcal M$ and any $i \in \mathcal M / \mathcal M_2$, we have $\mathsf{F}(\mathcal M'_1 \bigcup \{i\}) - \mathsf{F}(\mathcal M'_1) \geq \mathsf{F}(\mathcal M'_2 \bigcup \{i\}) - \mathsf{F}(\mathcal M'_2)$.
      \end{itemize}
    \end{definition}
    \begin{lemma} \label{le:submodularpro}
      Given any two disjoint subset $\mathcal M_1 \subseteq \mathcal M$ and $\mathcal M_2 \subseteq \mathcal M$ such that $\mathcal M_1 \bigcap \mathcal M_2 = \emptyset$, we have
      \begin{align} \label{eq:le:submodularpro}
        \sum_{i \in \mathcal M_1} \left( \mathsf{F}(\{i\} \bigcup \mathcal M_1) - \mathsf{F}(\mathcal M_1)\right) \geq \mathsf{F}(\mathcal M_1 \bigcup \mathcal M_2) - \mathsf{F}(\mathcal M_2)
      \end{align}
      if $\mathsf{F}(\cdot)$ is a monotone submodular set function.
    \end{lemma}
    \begin{proof}
      Supposing $\mathcal M_1 = \{i_1, i_2, \cdots, i_{M_1}\}$ and $\mathcal M_{1,\ell} = \{i_1, \cdots, i_{\ell}\}$, we have
      \begin{align} \label{eq:le:submodularpro00}
        &\mathsf{F}(\mathcal M_1 \bigcup \mathcal M_2)  \nonumber\\
        &= \mathsf{F}(\mathcal M_2) + \sum^{M_1}_{\ell=1} (\mathsf{F}(\mathcal M_2 \bigcup \mathcal M_{1,\ell}) - \mathsf{F}(\mathcal M_2 \bigcup \mathcal M_{1,\ell-1}))
      \end{align}
      When $\mathsf{F}(\cdot)$ is a monotone submodular set fucntion, we have $\mathsf{F}(\mathcal M_2 \bigcup \{i_\ell\}) \geq \mathsf{F}(\mathcal M_2 \bigcup \mathcal M_{1,\ell})$ and $\mathsf{F}(\mathcal M_2 \bigcup \mathcal M_{1,\ell}) \geq \mathsf{F}(\mathcal M_{2})$ for any $\ell =1,\cdots,M_1$. The proof can be completed by substituting the above inequalities into (\ref{eq:le:submodularpro00}).
    \end{proof}

\section{Our Proposed Method} \label{sec:metd}
  In this section, we introduce the details of our PDSR method. We consider $M$ services and $N$ users across $R$ different platforms. Let $\mathcal M=\{1,2,\cdots,M\}$, $\mathcal N=\{1,2,\cdots,N\}$ and $\mathcal R=\{1,2,\cdots,R\}$ denote the set of services, the set of users and the set of platforms, respectively. Suppose $\mathcal{N}_r \subseteq \mathcal N$ denotes the set of users on platform $r$, and $N_r$ is the cardinality of $\mathcal{N}_r$, i.e., the number of users on platform $r$. For each service $i \in \mathcal M$, it can be invoked by user $j\in\mathcal N_r$ through platform $r\in \mathcal R$. The resulting QoS data $d_{i,j,r} \geq 0$ is then observed by platform $r$ \footnote{We hereby deliberately assume that QoS values are represented by a non-negative real numbers. It could be user rating, response time, price, reliability of service, or even their combinations. As will be shown later, our method can be readily extended to handle multi-dimensional QoS data, e.g., by considering the similarity between matrices.}. Therefore, for each service $i$, its QoS value over the $N_r$ users on platform $r$ can be represented by a $N_r$-dimensional vector $\vec{d}_{i,r} = \left( d_{i,j,r} \right)_{j\in \mathcal N_r}$. Without loss of generality, we assume $d_{i,j,r} = 0$ if service $i$ is never invoked by user $j$ on platform $r$; otherwise, $d_{i,j,r} > 0$.
  
  Before diving into the details of our proposed method, we first introduce its outline. As demonstrated in Sec.~\ref{ssec:indexing}, each platform, e.g., $r$, first calculates a LSH vector for each of the services $i\in \mathcal M$ and then shares the vectors with each other. By concatenating the received vectors of any service $i \in \mathcal M$, platform $r$ can calculate an index $v_i$ for $\forall i$. The above procedure can be called repeatedly such that each platform $r$ can construct a similarity graph $\mathcal G$ where any two similar services are connected by an edge, as illustrated in Sec.~\ref{ssec:congraph}. When some particular platform $r^\dagger$ intends to recommend services to user $j^\dagger$, it seeks for a subset of $K$ services by maximizing our accuracy-diversity measure over the similarity graph $\mathcal G$.

  \begin{table*}
  \caption{Frequently Used Notations} \label{tab:notation}
  \begin{tabular}[t]{|p{5.8cm}|p{11cm}|}
  \hline
    $\mathcal M = \{1,2,\cdots,M\}$ & A set of $M$ services \\ \hline
    $\mathcal N = \{1,2,\cdots,N\}$ & A set of $N$ users \\ \hline
    $\mathcal R = \{1,2,\cdots,R\}$ & A set of $R$ platforms \\ \hline
    $\mathcal N_r \subset \mathcal N$ & A set of users on platform $r$ \\ \hline
    $N_r = |\mathcal N_r|$ & The number of users on platform $r$ \\ \hline
    ${d}_{i,j,r} \geq 0 $  & QoS value of service $i$ for user $j$ on platform $r$  \\ \hline
    $\vec{d}_{i,r} = \left( \vec{d}_{i,j,r} \right)_{j \in \mathcal N}$  & QoS vector of service $i$ on platform $r$  \\ \hline
    $\mathcal D = \left\{ \vec{d}_{i,r} \right\}_{i \in \mathcal N, r \in \mathcal R}$  & QoS dataset  \\ \hline
    $\mathcal{H}_{r} = \{ h_{r,1}, \cdots, h_{r,H_r} \}$ & A family of $H_r$ LSH functions on platform $r$    \\ \hline
    $\mathcal{H}_r \left( \vec{d}_{i,r} \right) = \left( h_{r,1}\left( \vec{d}_{i,r} \right), \cdots, h_{r,H_r}\left( \vec{d}_{i,r} \right) \right)$ & LSH vector of service $i$ on platform $r$  \\ \hline
    $v_i = \left( \mathcal{H}_1 \left( \vec{d}_{i,1} \right), \cdots, \mathcal{H}_R \left( \vec{d}_{i,R} \right) \right)$ & LSH index of service $i$\\ \hline
    $\mathcal{G} = (\mathcal{M}, \mathcal{E}) $ & Similarity graph $\mathcal G$ consisting of vertices $\mathcal{M}$ and edges $\mathcal E$ \\ \hline
    $j^\dagger, r^\dagger$ & Target user $j^\dagger$ and target platform $r^\dagger$\\ \hline
    $\overline{\mathcal M}_{j^\dagger} = \{i\in\mathcal M \mid d_{i,j^\dagger,r^\dagger}=0\}$ & Candidate services which we can recommend to user $j^\dagger$ on platform $r^\dagger$\\ \hline
    $\mathcal M_i = \{i' \in \mathcal M \mid (i,i') \in \mathcal E, d_{i',j^\dagger,r^\dagger} \neq 0\}$ & Services adjacent to $i$ in graph $\mathcal G$ which have been rated by user $j^\dagger$ on platform $r^\dagger$  \\ \hline
    \end{tabular}
  \end{table*}

  \subsection{Step 1: Indexing Services Based on LSH} \label{ssec:indexing}
    In this paper, we use \textit{cosine similarity} to construct our LSH mechanism \cite{Charikar-STOC02}. Specifically, given two points (or vectors) $v$ and $v'$ in a $N$-dimensional space, their cosine similarity can be defined by the cosine of the angle between them, i.e.,
    \begin{equation} \label{eq:cosine}
      \cos{\theta_{v,v'}} = \frac{v \cdot v'}{\|v\| \|v'\|}
    \end{equation}
    where $\theta_{v,v'} \in [-\pi, \pi]$ denotes the angle between $v$ and $v'$. Cosine similarity $\cos{\theta_{v,v'}} \in [-1,1]$ is a signed metric to measure the similarity between $v$ and $v'$. The larger is $\cos{\theta_{v,v'}}$, the smaller is the angle between $v$ and $v'$ and thus the more similar are $v$ and $v'$. 
  
    Considering a hyperplane in the $N$-dimensional space which passes the origin and has its normal vector defined by $v^*$, it divides the space into two half-space, i.e., positive half-space and negative half-space . We denote by $h_{v^*}(v): \mathbb R^N \rightarrow \{1, 0\}$ an indicator function parameterized by normal vector $v^*$, to specify the half-space containing point $v$. Specifically,
    \begin{equation} \label{eq:hyper}
      h_{v^*}(v) = \begin{cases}
        1, ~\cos{\theta_{v,v^*}} \geq 0 \\
        0, ~\text{otherwise}
      \end{cases}
    \end{equation}
    In another word, $v$ is in the positive half-space such that $h_{v^*}(v)=1$, if $v$ is ``similar'' to $v^*$ with $\cos{\theta_{v,v^*}} \geq 0$ (and thus $-\frac{\pi}{2} \leq \theta_{v,v^*} \leq \frac{\pi}{2}$); otherwise, $h_{v^*}(v)=0$. Given a random hyperplane defined by a random normal vector $v^*$ and two arbitrary points $v$ and $v'$, we have 
    \begin{equation}
      \mathbb{P}(h_{v^*}(v)=h_{v^*}(v')) = 1-\frac{|\theta_{v,v'}|}{\pi}
    \end{equation}
    Therefore, all functions $\{h_{v^*} \mid v^*\in[-\pi, \pi]\}$ composes a LSH function family.
  
    Recall that, for each service $i$, its quality on platform $r$ can be represented by a $N_r$-dimensional vector $\vec d_{i,r}$, where $N_r$ is the number of users on platform $r$. As shown in \textbf{Algorithm}~\ref{alg:servindex}, each platform $r$ randomly choose $H_r$ $N_r$-dimensional (normal) vectors to construct its $H_r$ LSH functions, i.e., $\mathcal H_r = \{h_{r,1}, h_{r,2}, \cdots, h_{r,H_r}\}$ (see Lin~\ref{ln:lshcons}). Based on the family of LSH functions $\mathcal H_r$, the hash values of each service $i$ are calculated under each of the hash functions and the hash values compose a $H_r$-dimensional binary vector 
    \begin{equation*}
      \mathcal H_r \left( \vec d_{i,r} \right) = \left( h_{r,1}\left(\vec d_{i,r}\right), h_{r,2}\left(\vec d_{i,r}\right), \cdots, h_{r,H_r}\left(\vec d_{i,r}\right) \right)
    \end{equation*}
    as shown in Lines 3$\sim$5. For each service $i\in\mathcal M$, it is indexed by concatenating the above vectors across different platforms. Specifically, as demonstrated in Line~7, the index of service $i$ can be represented by
    \begin{equation} \label{eq:index}
      v_i = \left( \mathcal H_1\left(\vec d_{i,1}\right), \mathcal H_2\left(\vec d_{i,2}\right), \cdots, \mathcal H_R\left(\vec d_{i,R}\right) \right)
    \end{equation}
    Note that $v_i \in \{0,1\}^{\sum^R_{r=1}H_r}$ for any service $i$.
    \begin{algorithm}[htb!]
      \KwIn{Platforms $\mathcal R$, services $\mathcal M$, users $\mathcal N$, QoS data $\mathcal D = \left\{\vec d_{i,r}\right\}_{i\in\mathcal M, r\in\mathcal R}$, the number of hash functions $H_r$ for any platform $r\in \mathcal R$.}
      \KwOut {Similarity index $v_i$ for any $i \in \mathcal M$.}
      \For{$r=1,2,\cdots,R$}{ 
        Choose $H_r$ random normal vectors and construct a family of LSH functions $\mathcal H_r$;  \label{ln:lshcons} \\
        \For{$i=1,2,\cdots,M$}{ 
          %
          %
          Calculate a LSH vector $\mathcal H_r \left(\vec d_{i,r}\right)$; \label{ln:lshvec}\\
        }
      }
      Calculate the index $v_i$ for service $i$, by concatenating the LSH vectors; \label{ln:concvec} 
    \caption{A LSH-based algorithm for service indexing.}
    \label{alg:servindex}
    \end{algorithm}

  \subsection{Step 2: Constructing Service Similarity Graph} \label{ssec:congraph}
    In the above step, we construct a cosine similarity-based LSH mechanism, by which each service is mapped to a binary hash value such that it is highly likely for similar services to have the same hash value. Therefore, we can measure the similarity among the services according to their hash values, so as to serve the goal of privacy preservation as mentioned in Sec.~\ref{ssec:lsh}. 

    In this step, we construct a graph $\mathcal G = (\mathcal M, \mathcal E)$ to characterize the similarity among the services based on their LSH vectors. The graph contains $\mathcal M$ as vertices and we propose \textbf{Algorithm}~\ref{alg:graph} to add edges $\mathcal E$ to the graph according to the similarity among the vertices (i.e., the services). As shown in Line 1, $\mathcal E$ is first initialized by $\mathcal E \leftarrow \emptyset$. In each iteration $t=1,\cdots,T$ of \textbf{Algorithm}~\ref{alg:graph}, \textbf{Algorithm}~\ref{alg:servindex} is called to create a LSH table $\mathsf{Table}_t = \{(i, v_i)\}_{i\in\mathcal{M}}$ (see Line~\ref{ln:contab}). The table is composed of $M$ entries and each entry is a tuple of $(i, v_i)$ . Then, for each pair of services $i$ and $i'$, if $(i,i') \notin \mathcal{E}$ and $v_i = v_{i'}$ (see Lines \ref{ln:addedge-start}-\ref{ln:addedge-end}). It is demonstrated that, any two services $i$ and $i'$ is connected in the similarity graph if they have the same LSH indices in any one of the $T$ iterations.
    \begin{algorithm}
    \KwIn{Platforms $\mathcal R$, services $ \mathcal M$, users $\mathcal N$, the number of iterations $T$.} 
    \KwOut{Service graph $\mathcal G = (\mathcal M, \mathcal E)$.}
      $\mathcal E = \emptyset$; \\
      \For {$t=1,\cdots,T$}{
        Call \textbf{Algorithm}~\ref{alg:servindex} to calculate the LSH table $\mathsf{Table}_t = \{(i, v_i)\}_{i \in \mathcal{M}}$; \label{ln:contab}\\
        \For{$(i,i') \in \mathcal M \times \mathcal M$}{ \label{ln:addedge-start}
          \If{$(v_i = v_{i'}) \&\& ((i,i') \notin \mathcal E)$}{
          $\mathcal E \leftarrow \mathcal E \bigcup (i,i')$;  \label{ln:adedge}\\
          } 
        }\label{ln:addedge-end}
      }
    \caption{Our method of constructing a weighted service similarity graph.}
    \label{alg:graph}
    \end{algorithm}

    As shown in \textbf{Algorithm}~\ref{alg:servindex}, for each service, its index is calculated by concatenating its LSH vectors across the different platform. The service indexing algorithm is then called iteratively to construct a similarity graph in \textbf{Algorithm}~\ref{alg:graph}. According to the definition of LSH mechanism (see \textbf{Definition}~\ref{def:lsh}), similar services have the same LSH indices with high probability and will be connected in the similarity graph. In another word, each platform can construct a service similarity graph without sharing the original data with each other, while the privacy preservation can be ensured thanks to the irreversibility of the hash functions. Furthermore, as shown in \textbf{Algorithm}~\ref{alg:servindex} and \textbf{Algorithm}~\ref{alg:graph}, the numbers of hash functions $\{H_r\}_{r\in \mathcal R}$ and the number of hash tables $T$ both impact the constructions of the similarity graph. Their effect is described in \textbf{Theorem}~\ref{thm:eff}.
    \begin{theorem} \label{thm:eff}
      Suppose $\theta_{i,i',r}$ represents the angle between $\vec{d}_{i,r}$ and $\vec{d}_{i',r}$ for any service $i,i' \in \mathcal M$. The probability for $i$ and $i'$ to have an edge in $\mathcal G$ can be written as
      \begin{equation} \label{eq:eff}
        \mathbb P((i,i') \in \mathcal E) = 1 - \left( 1- \prod^R_{r=1} \left( 1- \frac{|\theta_{i,i',r}|}{\pi}  \right)^{H_r}\right)^T
      \end{equation}
    \end{theorem}
    \begin{proof}
      For any services $i$ and $i'$ on platform $r$, suppose $\theta_{i,i',r} \in \left[ -\frac{\pi}{2}, \frac{\pi}{2} \right]$ denotes the angle between $\vec{d}_{i,r}$ and $\vec{d}_{i',r}$. Assume $X_{i,i',\ell} \in \{0,1\}$ be a Bernoulli random variable indicating if $i$ and $i'$ have the same hash value under the $\ell$-th hash function; we then have $\mathbb{P}(X_{i,i',\ell}=1) = 1-\frac{|\theta_{i,i',r}|}{\pi}$ and $\mathbb{P}(X_{i,i',\ell}=0) = \frac{|\theta_{i,i',r}|}{\pi}$. Since each platform chooses their hash functions independently, when there are $H_r$ hash functions on platform $r$, the probability of $v_i = v_{i'}$ can be written as
      \begin{equation}
        \mathbb P(v_i = v_{i'}) = \prod^R_{r=1} \left( 1 - \frac{|\theta_{i,i',r}|}{\pi}  \right)^{H_r}
      \end{equation}
      According to \textbf{Algorithm}~\ref{alg:graph}, we perform the above similarity calculation $T$ times, and add an edge between $i$ and $i'$ if we judge they are similar to each other once. Hence, we have $(i,i') \in \mathcal E$ with probability (\ref{eq:eff})
    \end{proof}

    It is implied by the above theorem that $H_r$ and $T$ actually determine the ``resolution'' at which we make our similarity judgement and thus impact our QoS prediction. If we take very small values for $H_r$ and $T$, some dissimilar services (with large angles between their QoS vector) may be thought of as similar ones with the same index; while when the values of $H_r$ and $T$ are large, we may assign different index values to those similar services. We will report some empirical results in Sec.~\ref{sec:experiment} to show the effect of $H_r$ and $T$.

  \subsection{Step 3: Recommending with Diversity} \label{ssec:recomm}

    Now we have graph $\mathcal G$ where similar vertices (i.e., services) are connected by edges. We assume user $j^\dagger \in \mathcal N$ is the one to whom we intend to recommend new services (i.e., the services user $j^\dagger$ never invoked) through platform $r^\dagger$. Let $\overline{\mathcal M}_{j^\dagger} = \{i\in\mathcal M \mid d_{i,j^\dagger,r^\dagger}=0\}$ be the set of target services. We first need to estimate the quality of these services. In particular, for any service $i \in \overline{\mathcal M}_{j^\dagger}$, we define $\mathcal M_i = \{i' \in \mathcal M \mid (i,i') \in \mathcal E, d_{i',j^\dagger,r^\dagger} \neq 0\}$ as the set of services which have been invoked by user $j^\dagger$ on platform $r^\dagger$ and are similar to service $i$. 
    By exploiting the similarity graph, we then predict the quality of any service $i \in \overline{\mathcal M}_{j^\dagger}$ as follows      
    \begin{equation} \label{eq:qosest}
      d_{i, j^\dagger, r^\dagger} = \frac{1}{M_i} \sum_{i' \in \mathcal M_i} d_{i',j^\dagger,r^\dagger}
    \end{equation}
    where $M_i = |\mathcal M_i|$ denotes the cardinality of $\mathcal M_i$, i.e., the number of services in $\mathcal M_i$.

    According to the predicted QoS values, one may want to recommend the services with the highest predicted QoS values to the target user; however, such a trivial choice may results in a homogeneous recommendation list which may not match the user's interests, as mentioned in Sec.~\ref{sec:intro}. Therefore, we are now interest in seeking for a trade-off between accuracy and diversity. In the following, we first present our  design of accuracy-diversity metric and formulate our $K$-Diversified Service Recommendation ($K$-DSR) problem in Sec.~\ref{sssec:form}. Since the $K$-DSR problem is NP-hard (see \textbf{Theorem}~\ref{thm:nphard}), we propose a greedy approximation algorithm in Sec.~\ref{sssec:algo} and prove its approximation ratio in Sec.~\ref{sssec:analysis}.

    \subsubsection{Problem Formulation} \label{sssec:form}
      Given any subset $\mathcal K \subseteq \mathcal M$, let $\mathsf{Acc}(\mathcal K)$ and $\mathsf{Div}(\mathcal K)$ be two real valued set functions to measure the accuracy and diversity of $\mathcal K$. Our $K$-Diversified Service Recommendation (K-DSR) problem then can be formulated as 
      \begin{align}
        &\max_{\mathcal K \subseteq \overline{\mathcal M}_{j^\dagger}} ~~ \mathsf{F}(\mathcal K) =  \mathsf{Acc}(\mathcal K) + \lambda \mathsf{Div}(\mathcal K)  \label{eq:obj} \\
        &~~~~\mathrm{s.t.} ~~~~~~~~ |\mathcal K| = K \label{eq:cons}
      \end{align}
      where $K$ is the number of services we intend to recommend and $\lambda$ denotes a factor to make a trade-off between recommendation accuracy and diversity \footnote{It is worthy to note that we hereby formulate the optimization for non-repetitive recommendation problems where we recommend the services which are never invoked by the target users, while our proposed method actually can be applied to the case where a service can be recommended repetitively.}. Therein, as shown in Eq.~(\ref{eq:accuracy}), we define set function $\mathsf{Acc}(\mathcal K)$ to recommend services with high estimated QoS values to the target users, so as to ensure the recommendation accuracy. 
      \begin{equation} \label{eq:accuracy}
        \mathsf{Acc}(\mathcal K) = \sum_{i\in\mathcal K} d_{i,j^\dagger,r^\dagger} 
      \end{equation}
      As for the recommendation diversity $\mathsf{Div}: 2^{\overline{\mathcal M}_{j^\dagger}} \rightarrow \mathbb R$, we take into account two types of diversity, i.e., \textit{direct} diversity and \textit{indirect} diversity. For any subset $\mathcal K \subseteq \overline{\mathcal M}_{j^\dagger}$, we design a metric to measure its direct diversity based on \textit{Jaccard dissimilarity}. Specifically, given two services $i$ and $i'$ as well as their QoS data $\vec{d}_{i,r^\dagger} = (d_{i,1,r^\dagger}, \cdots, d_{i,N,r^\dagger})$ and $\vec{d}_{i',r^\dagger} = (d_{i',1,r^\dagger}, \cdots, d_{i',N,r^\dagger})$ which are predicted upon the similar graph, their \textit{Jaccard dissimilarity} can be calculated by
      \begin{align} \label{eq:jaccard}
        &\mathsf{J}\left(\vec{d}_{i,r^\dagger},\vec{d}_{i',r^\dagger}\right) \nonumber\\
        =& 1-\frac{\sum^N_{j=1} \mathbb{I}((d_{i,j,r^\dagger} \neq 0) \land (d_{i',j,r^\dagger} \neq 0))}{N - \sum^N_{j=1} \mathbb{I}((d_{i,j,r^\dagger}=0) \land (d_{i',j,r^\dagger}=0))}
      \end{align}
      where $\mathbb{I}: \{True, False\} \rightarrow \{1, 0\}$ is an indicator function.
      The direct diversity of $\mathcal K$ then is defined as
      \begin{equation} \label{eq:dirdiv}
        \beta(\mathcal K) = \sum_{i,i'\in \mathcal K} \mathsf{J} \left( \vec{d}_{i,r^\dagger}, \vec{d}_{i',r^\dagger} \right)
      \end{equation}
      The subset $\mathcal K$ is with higher direct diversity if the services in the subset are less similar to each other. Nevertheless, considering the direct diversity only is not sufficient, since the direct diversity is calculated based on the estimates on the qualities of the services, which inevitably induces bias. Therefore, in addition to the direct diversity, we define a notion of indirect diversity as a complement to correct the bias. According to what we have mentioned in Sec.~\ref{ssec:expansion}, it is more likely for the services in $\mathcal K$ to share the same expanded neighbors in graph $\mathcal G$ when $\mathcal K$ has smaller expansion ratio, while the service may be similar to each other if they are connected to and thus similar to the same expanded neighbors in graph $\mathcal G$. Hence, for any subset $\mathcal K \subseteq \mathcal M$, it is with higher (indirect) diversity if it has larger expansion ratio $\alpha(\mathcal K)$. By combining both the direct and indirect diversities, we define
      \begin{equation} \label{eq:div}
        \mathsf{Div}(\mathcal K) =  \alpha(\mathcal K) + \xi \beta(\mathcal K)
      \end{equation}
      where $\xi>0$ is a weight factor such that we can fine tune the trade-off between the direct and indirect diversity. 
    
      \begin{theorem} \label{thm:nphard}
        Our $K$-DSR problem defined by (\ref{eq:obj}) and (\ref{eq:cons}) is NP-hard.
      \end{theorem}
      \begin{proof}
        The basic idea of the proof is to show the NP-harness of the following two sub-problems
        \begin{align*}
          \mathfrak{P}_1:~~&\max_{\mathcal K \subseteq \overline{\mathcal M}_{j^\dagger}} ~~\alpha(\mathcal K)  \\
          &~~~~\mathrm{s.t.} ~~~~~~~~|\mathcal K| = K
        \end{align*}
        and
        \begin{align*}
          \mathfrak{P}_2:~~&\max_{\mathcal K \subseteq \overline{\mathcal M}_{j^\dagger}} ~~\beta(\mathcal K)  \\
          &~~~~\mathrm{s.t.} ~~~~~~~~|\mathcal K| = K
        \end{align*}
        The first problem $\mathfrak{P}_1$ is to find a subset $\mathcal K \subseteq \overline{\mathcal M}_{j^\dagger}$ of size $K$ with the maximum expansion ratio. We prove its NP-hardness by demonstrating a reduction from the maximum coverage problem (which is a well-known NP-hard problem). In the maximum coverage problem, we assume $\mathcal M'$ denotes a set of $M'$ elements and $\mathcal C = \{\mathcal C_1, \mathcal C_2, \cdots, \mathcal C_C\}$ represent a collection of $C$ subsets of $\mathcal M'$ such that $\bigcup_\ell \mathcal C_\ell = \mathcal M'$. Given a positive integer $K < C$, the goal of the maximum coverage problem is to find $K$ out of the $C$ subsets such that their union has the maximum cardinality. To construct an instance of the maximum coverage problem, for each $\mathcal C_\ell \in \mathcal C$ and $i \in \mathcal M'$, we create node $a_\ell$ and $b_i$, respectively. If $i \in \mathcal C_\ell$, we add a directed edge from $a_\ell$ to $b_i$ and it is said that $b_i$ (resp. element $i \in \mathcal M'$) is ``covered'' by $a_\ell$ (resp. subset $\mathcal C_\ell$). Let $\mathcal A = \{a_\ell\}_{\ell}$ and $\mathcal B = \{b_i\}_{i}$. The edges are only from the nodes in $\mathcal A$ to the ones in $\mathcal B$; then, for any feasible solution to the maximum coverage problem $\widetilde{\mathcal C} \subseteq \mathcal C$, its expansion ratio is calculated by $\left|\mathcal B_{\widetilde{\mathcal C}}\right| / K$, where $\mathcal B_{\widetilde{\mathcal C}} \subseteq \mathcal B$ denotes the subset of the nodes in $\mathcal B$ covered by $\widetilde{\mathcal C}$, i.e., the union of the subsets in $\widetilde{\mathcal C}$. Therefore, finding a subset of $K$ nodes in $\mathcal A$ with the highest expansion ratio is equivalent to finding $K$ subsets in $\mathcal C$ with the highest coverage, and vice versa. 

        As for the second optimization problem $\mathfrak{P}_2$, it is actually a maximum dispersion problem, since the Jaccard distance (or dissimilarity) satisfies triangle inequality~\cite{HassinRT-ORL97}. Therefore, the NP hardness of our problem can be proved.
      \end{proof}

    \subsubsection{Algorithm Design} \label{sssec:algo}
      Due to the NP-hardness of the optimization problem, we have to be content with approximation algorithms. We propose a greedy algorithm to address the problem in this section, as shown in \textbf{Algorithm}~\ref{alg:greedy}. Let $\mathcal K$ denote the recommended list which is initialized as an empty set (see Line~\ref{ln:initK}). The algorithm proceeds iteratively. In each iteration, as shown in Line~\ref{ln:find}, we first find service $i^\dagger \in \overline{\mathcal M}_{j^\dagger}$ such that the set function $\mathsf{F}'(\mathcal K \cup \{i^\dagger\})$ can be maximized, where $\mathsf{F}': 2^{\overline{\mathcal M}_{j^\dagger}} \rightarrow \mathbb{R}$ is defined as
      \begin{align} \label{eq:fprime}
        \mathsf{F}'(\mathcal K) = \frac{1}{2} \left( \mathsf{Acc}(\mathcal K) +  \lambda\alpha(\mathcal K) \right) + \lambda\xi \beta(\mathcal K)
      \end{align}
      for $\forall \mathcal{K} \subseteq \overline{\mathcal M}_{j^\dagger}$. We then add $i^\dagger$ into $\mathcal K$ and remove $i^\dagger$ from $\overline{\mathcal M}_{j^\dagger}$, as demonstrated in Line~\ref{ln:add} and Line~\ref{ln:remove}, respectively.

      \begin{algorithm}[htb!]
        \KwIn{Service similarity graph $\mathcal G = (\mathcal M, \mathcal E)$, target user  $j^\dagger$ and target platform $r^\dagger$, QoS data $\mathcal D$, the number of recommended entries $K$.}
        \KwOut{Recommended list $\mathcal K \subseteq \mathcal M$ for user $j^\dagger$ on platform $r^\dagger$.}

        $\mathcal K= \emptyset$; \label{ln:initK}\\
        \While {$\lvert \mathcal K \rvert < K $ }{ \label{ln:loopstart}
          $i^\dagger = \arg\max_{i\in \overline{\mathcal M}_{j^\dagger}} \mathsf{F}'(\mathcal K \bigcup \{i\})$; \label{ln:find}\\
          $\mathcal K \leftarrow \mathcal K \cup \{i^\dagger\}$; \label{ln:add}\\
          $\overline{\mathcal M}_{j^\dagger} \leftarrow \overline{\mathcal M}_{j^\dagger} / \{i^\dagger\}$; \label{ln:remove}\\
        } \label{ln:loopend}
      \caption{Top-$K$ service recommendation algorithm}
      \label{alg:greedy}
      \end{algorithm}

    \subsubsection{Analysis} \label{sssec:analysis}
      In this section, we will analyze the performance of \textbf{Algorithm}~\ref{alg:greedy}. Before showing its approximation ratio in \textbf{Theorem}~\ref{thm:appratio}, we first present two lemmas as follows, which will be very helpful in the proof of \textbf{Theorem}~\ref{thm:appratio}. 
      \begin{lemma} \label{le:submodu}
        For $\forall \mathcal K \subseteq \mathcal M$, the following set function defined by 
        \begin{eqnarray} \label{eq:fpri}
          \mathsf{\Phi}'(\mathcal K) = \mathsf{Acc}(\mathcal K) + \lambda \alpha(\mathcal K) =  \sum_{i\in \mathcal K} d_{i,j^\dagger,r^\dagger} + \frac{\lambda \lvert \mathsf{Exp(\mathcal K)} \rvert}{M} 
        \end{eqnarray}
        is a monotone submodular set function.
      \end{lemma}
      \begin{proof}
        For $\forall \mathcal K_1 \subseteq \mathcal K_2 \subseteq \mathcal M$, we have $\mathsf{Exp}(\mathcal K_1) \subseteq \mathsf{Exp}(\mathcal K_2)$ and $\sum_{i \in \mathcal K_1} d_{i,j^\dagger,r^\dagger} \leq \sum_{i \in \mathcal K_2} d_{i,j^\dagger,r^\dagger}$. Furthermore, $\forall i \in \mathcal M / \mathcal K_2$, we have
        \begin{align}
          &\mathsf{\Phi}'(\mathcal K_1 \cup \{i\}) - \mathsf{\Phi}'(\mathcal K_1) - \left( \mathsf{\Phi}'(\mathcal K_2 \cup \{i\}) - \mathsf{\Phi}'(\mathcal K_2) \right)  \nonumber\\
          =& \lambda \frac{|\mathsf{Exp}(\mathcal{K}_1 \cup \{i\})|-|\mathsf{Exp}(\mathcal{K}_1)|}{M} \nonumber\\
          &- \lambda\frac{|\mathsf{Exp}(\mathcal{K}_2 \cup \{i\})|-|\mathsf{Exp}(\mathcal{K}_2)|}{M} 
        \end{align}
        Therein, $|\mathsf{Exp}(\mathcal{K}_1 \bigcup \{i\})|-|\mathsf{Exp}(\mathcal{K}_1)| = |\mathsf{Exp}(\{i\}) -  \mathsf{Exp}(\mathcal{K}_1)|$ and $|\mathsf{Exp}(\mathcal{K}_2 \bigcup \{i\})|-|\mathsf{Exp}(\mathcal{K}_2)| = |\mathsf{Exp}(\{i\}) -  \mathsf{Exp}(\mathcal{K}_2)|$; we then obtain
        \begin{align}
          &\mathsf{\Phi}'(\mathcal K_1 \cup \{i\}) - \mathsf{\Phi}'(\mathcal K_1) - \left( \mathsf{\Phi}'(\mathcal K_2 \cup \{i\}) - \mathsf{\Phi}'(\mathcal K_2) \right)  \nonumber\\
          =& \frac{\lambda ( |\mathsf{Exp}(\{i\}) -  \mathsf{Exp}(\mathcal{K}_1)| - |\mathsf{Exp}(\{i\}) -  \mathsf{Exp}(\mathcal{K}_2)| )}{M} \geq 0
        \end{align}
        which completes the proof according to \textbf{Definition}~\ref{def:modufunc}.
      \end{proof}
      \begin{lemma} \label{le:inequality}
        Given the \textit{Jaccard dissimilarity} $\mathsf{J(\cdot,\cdot)}$ defined in Eq.~(\ref{eq:jaccard}), for any two disjoint nonempty sets $\mathcal K_1, \mathcal K_2 \subseteq \mathcal M$, we have the following inequality:
        \begin{eqnarray} \label{eq:ineq}
          (\lvert \mathcal K_1 \rvert - 1) \mathsf{\Gamma}(\mathcal K_1,\mathcal K_2)
          \geq \lvert \mathcal K_2 \rvert \mathsf{\Gamma}(\mathcal K_1)
        \end{eqnarray}
        where $\mathsf{\Gamma}(\mathcal K_1) =  \sum_{i, i'\in \mathcal K_1 } \mathsf{J}(d_{i,j^\dagger, r^\dagger},d_{i',j^\dagger, r^\dagger})$ and $\mathsf{\Gamma}(\mathcal K_1,\mathcal K_2) = \sum_{i \in \mathcal K_1} \sum_{i'\in \mathcal K_2} \mathsf{J}(d_{i, j^\dagger, r^\dagger}, d_{i', j^\dagger, r^\dagger})$, 
      \end{lemma}
      \begin{proof}
        For any $i_2 \in \mathcal K_2$ and any $i_1, i'_1 \in \mathcal K_1$ such that $i_1 \neq i'_1$, we have
        \begin{align}
          &\mathsf{J}(d_{i_2,j^\dagger,r^\dagger}, d_{i_1,j^\dagger,r^\dagger}) + \mathsf{J}(d_{i_2,j^\dagger,r^\dagger}, d_{i'_1,j^\dagger,r^\dagger})  \nonumber\\
          &\geq \mathsf{J}(d_{i_1,j^\dagger,r^\dagger}, d_{i'_1,j^\dagger,r^\dagger})
        \end{align}
        according to the triangle inequality; we thus obtain 
        \begin{align}
          \mathsf{\Gamma}(\mathcal K_1) \leq& \mathsf{J}(d_{i_1,j^\dagger,r^\dagger}, d_{i'_1,j^\dagger,r^\dagger})  \nonumber\\ 
          =& (|\mathcal K_1|-1) \sum_{i_1\in\mathcal K_1} \mathsf{J}(d_{i_1,j^\dagger,r^\dagger}, d_{i_2,j^\dagger,r^\dagger})
        \end{align}
        for any $i_2 \in \mathcal K_2$. The lemma can be completed by summing the above inequality over $i_2 \in \mathcal K_2$.
      \end{proof}
      Based on the \textbf{Definition}~\ref{def:modufunc}, the problem defined in Eq.(8) and Eq.(9) can be regarded as the max-sum diversification problem with monotone submodular set functions satisfying a cardinality constraint. Thus, \textbf{Theorem}~\ref{thm:appratio} is as follows.

      \begin{theorem} \label{thm:appratio}
        Our top-$K$ recommendation algorithm (see \textbf{Algorithm}~\ref{alg:greedy}) is a $2$-approximation algorithm.
      \end{theorem}
      \begin{proof}
        Let $\mathcal K^*$ be the optimal solution to the problem K-DSR. For any service $i \in \mathcal K^*$, it is said to be ``correct''; otherwise, it is ``incorrect''. Recall that $\mathcal K^\dagger$ denotes the recommendation list yielded by our greedy algorithm, and let $\mathcal K^\dagger_{[t]}$ be the (intermediate) output in the $t$-th iteration. Assume ${\mathcal K}_{[t]} = \mathcal K^* \bigcap \mathcal K^\dagger_{[t]}$, $\widetilde{\mathcal K}^\dagger_{[t]} = \mathcal{K}^\dagger_{[t]}/{\mathcal K}_{[t]}$ and $\mathcal{K}^*_{[t]} = \mathcal K^* / {\mathcal K}_{[t]}$. In another word, $\mathcal K_{[t]}$ and $\widetilde{\mathcal K}^\dagger_{[t]}$ denote the set of correct services and the set of incorrect ones we select till the $t$-th iteration, respectively, while $\mathcal{K}^*_{[t]}$ is the set of correct services we do not select till the $t$-th iteration. Since $\mathcal K_{[t]} \bigcup \mathcal{K}^*_{[t]} = \mathcal K^*$, it follows that
        \begin{align} \label{eq:appratio00}
          \mathsf{\Gamma} \left( \mathcal K_{[t]}, \mathcal{K}^*_{[t]} \right) + \mathsf{\Gamma} \left(\mathcal K_{[t]}\right) + \mathsf{\Gamma}\left(\mathcal{K}^*_{[t]}\right) - \mathsf{\Gamma}\left(\mathcal{K}^*\right) = 0
        \end{align}
        Furthermore, according to \textbf{Lemma}~\ref{le:inequality}, we have the following inequalities
        \begin{subequations} \label{eq:inequs}
        \begin{equation} \label{eq:inequs-a}
          \left(|\mathcal{K}^*_{[t]}|-1\right) \cdot \mathsf{\Gamma}\left(\widetilde{\mathcal K}^\dagger_{[t]}, \mathcal{K}^*_{[t]}\right) - |\widetilde{\mathcal K}^\dagger_{[t]}| \cdot \mathsf{\Gamma}\left(\mathcal{K}^*_{[t]}\right) \geq 0
        \end{equation}
        \begin{equation} \label{eq:inequs-b}
          \left(|\mathcal{K}^*_{[t]}|-1\right) \cdot \mathsf{\Gamma}\left(\mathcal K_{[t]}, \mathcal{K}^*_{[t]}\right) - |\mathcal K_{[t]}| \cdot \mathsf{\Gamma}\left(\mathcal{K}^*_{[t]}\right) \geq 0
        \end{equation}
        \begin{equation} \label{eq:inequs-c}
          \left(|\mathcal K_{[t]}|-1\right) \cdot \mathsf{\Gamma}\left(\mathcal K_{[t]}, \mathcal{K}^*_{[t]}\right) - |\mathcal{K}^*_{[t]}| \cdot \mathsf{\Gamma}\left(\mathcal K_{[t]}\right) \geq 0
        \end{equation}
        \end{subequations}

        Before diving into the details of the proof, we first introduce some symbols and notations. For any $\mathcal K \subseteq \overline{\mathcal M}_j$ and $i \in \overline{\mathcal M}_j / \mathcal K$, suppose
        \begin{align*}
        \begin{cases}
          \Delta\mathsf{\Phi}'_{i}(\mathcal K) = \mathsf{\Phi}'\left(\{i\} \bigcup \mathcal K\right) - \mathsf{\Phi}'\left(\mathcal K\right) \\
          \Delta{\beta}_i(\mathcal K) = \beta \left( \{i\} \bigcup \mathcal K \right) - \beta \left( \mathcal K \right) \\
          \Delta\mathsf{F}'_i(\mathcal K) = \mathsf{F}' \left( \{i\} \bigcup \mathcal K \right) - \mathsf{F}' \left( \mathcal K \right) = \frac{1}{2}\Delta\mathsf{\Phi}'_{i}(\mathcal K) + \lambda \xi \Delta{\beta}_i(\mathcal K)
        \end{cases}
        \end{align*}

        \textbf{Case}~1: We first take into account the first case where $| \mathcal{K}^*_{[t]} | = 1$ and thus $|\mathcal K^\dagger_{[t]}| = K-1$ and $\mathcal K^\dagger_{[t]} \subseteq \mathcal K^*$ in the $t$-th iteration. In another word, $K-1$ out of the $K$ correct services are selected by our greedy algorithm till the $t$-th iteration. We denote by $i^*$ the only element in $\mathcal{K}^*_{[t]}$. We assume $i_{[t+1]}$ is the service selected by our greedy algorithm in the following $(t+1)$-th iteration; we then have
        \begin{align}
          \Delta\mathsf{F}'_{i_{[t+1]}} \left(\mathcal{K}^\dagger_{[t]}\right) \geq&  \Delta\mathsf{F}'_{i^*} \left(\mathcal{K}^\dagger_{[t]}\right) \nonumber\\
          =& \frac{1}{2} \Delta \mathsf{\Phi}'_{i^*} \left(\mathcal{K}^\dagger_{[t]}\right) + \lambda\xi \Delta\beta_{i^*} \left(\mathcal{K}^\dagger_{[t]}\right) \nonumber\\
          \geq& \frac{1}{2} \left( \Delta \mathsf{\Phi}'_{i^*} \left(\mathcal{K}^\dagger_{[t]}\right) + \lambda\xi \Delta\beta_{i^*} \left(\mathcal{K}^\dagger_{[t]}\right) \right) \nonumber\\
          =& \frac{1}{2} \Delta \mathsf{F}_{i^*} \left( \mathcal{K}^\dagger_{[t]} \right)
        \end{align}
        and thus
        \begin{align}
          \Delta \mathsf{F}_{i_{[t+1]}}(\mathcal K^\dagger_{[t]}) =& \mathsf{F} \left( \left\{i_{[t+1]}\right\} \bigcup \mathcal K^\dagger_{[t]} \right) - \mathsf{F}\left(\mathcal{K}^\dagger_{[t]}\right) \nonumber\\
          =& \Delta\mathsf{\Phi}'_{i_{[t+1]}}(\mathcal K^\dagger_{[t]}) + \lambda \xi \Delta\beta_{i_{[t+1]}} (\mathcal K^\dagger_{[t]}) \nonumber\\
          \geq& \frac{1}{2} \Delta\mathsf{\Phi}'_{i_{[t+1]}}(\mathcal K^\dagger_{[t]}) + \lambda \xi \Delta\beta_{i_{[t+1]}} (\mathcal K^\dagger_{[t]})  \nonumber\\
          =& \Delta\mathsf{F}'_{i_{[t+1]}} (\mathcal K^\dagger_{[t]}) \nonumber\\
          \geq& \frac{1}{2} \Delta \mathsf{F}_{i^*}(\mathcal K^\dagger_{[t]})
        \end{align}
        according to which, we obtain
        \begin{align}
          \mathsf{F}\left(\mathcal{K}^\dagger\right) =& \mathsf{F}\left(\mathcal{K}^\dagger_{[t]}\right) + \Delta\mathsf{F}_{i_{[t+1]}}\left(\mathcal{K}^\dagger_{[t]}\right)  \nonumber\\
          \geq& \mathsf{F}\left(\mathcal{K}^\dagger_{[t]}\right) + \frac{1}{2} \Delta \mathsf{F}_{i^*}(\mathcal K^\dagger_{[t]})  \nonumber\\
          \geq& \frac{1}{2} \left( \mathsf{F}\left(\mathcal{K}^\dagger_{[t]}\right) + \Delta \mathsf{F}_{i^*}(\mathcal K^\dagger_{[t]}) \right)  \nonumber\\
          =& \frac{1}{2} \left( \mathsf{F}\left(\mathcal{K}^\dagger_{[t]}\right) + \mathsf{F}\left(\mathcal{K}^*\right) - \mathsf{F}\left(\mathcal{K}^\dagger_{[t]}\right) \right) \nonumber\\
          =& \frac{1}{2} \mathsf{F}\left(\mathcal{K}^*\right)
        \end{align}

        \textbf{Case 2}: We then consider the case with $|\mathcal K^*_{[t]}| >1$. Since $|\mathcal K^*_{[t]}| >1$, $K \geq 1$ and $|\mathcal K^*_{[t]}| - | \widetilde{\mathcal K}^\dagger_{[t]} | = K - t \geq 0$, we take them as multipliers and calculate
        \begin{align}
          &\frac{1}{|\mathcal{K}^*_{[t]}|-1} \cdot (\ref{eq:inequs-a}) + \frac{|\mathcal{K}^*_{[t]}|-|\widetilde{\mathcal K}^\dagger_{[t]}|}{K \left( |\mathcal{K}^*_{[t]}|-1 \right)} \cdot (\ref{eq:inequs-b})  \nonumber\\
          +& \frac{t}{K(K-1)} \cdot (\ref{eq:inequs-c}) + \frac{t|\mathcal{K}^*_{[t]}|}{K(K-1)} \cdot (\ref{eq:appratio00})
        \end{align}
        as
        \begin{align}
          &\gamma_0 \cdot \mathsf{\Gamma} \left( \mathcal{K}_{[t]}, \mathcal{K}^*_{[t]} \right) + \mathsf{\Gamma} \left( \widetilde{\mathcal K}^\dagger_{[t]}, \mathcal{K}^*_{[t]} \right)  \nonumber\\
          &+ \gamma_1 \cdot \mathsf{\Gamma}( \mathcal{K}^*_{[t]} ) - \frac{k|\mathcal{K}^*_{[t]}|}{K(K-1)} \cdot \mathsf{\Gamma}(\mathcal{K}^*) \geq 0
        \end{align}
        where
        \begin{align}
          \gamma_0 = \frac{|\mathcal{K}^*_{[t]}|-|\widetilde{\mathcal K}^\dagger_{[t]}|}{K} + \frac{k(|\mathcal{K}_{[t]}|+|\mathcal{K}^*_{[t]}|-1)}{K(K-1)} = 1
        \end{align}
        and
        \begin{align}
          \gamma_1 =& \frac{k|\mathcal{K}^*_{[t]}|}{K(K-1)} - \frac{|\mathcal{K}_{[t]}|(|\mathcal{K}^*_{[t]}|-|\widetilde{\mathcal K}^\dagger_{[t]}|)}{K(|\mathcal{K}^*_{[t]}|-1)} - \frac{|\widetilde{\mathcal K}^\dagger_{[t]}|}{|\mathcal{K}^*_{[t]}|-1} \nonumber\\
          =& \frac{k|\mathcal{K}^*_{[t]}|(K-|\mathcal{K}^*_{[t]}|)}{K(K-1)(|\mathcal{K}^*_{[t]}|-1)} \geq 0
        \end{align}
        as $|\mathcal{K}^*_{[t]}| = K - |\mathcal{K}_{[t]}|$, $|\mathcal{K}^*_{[t]}| = t = |\mathcal{K}_{[t]}|$ and $|\mathcal{K}^*_{[t]}| - |\widetilde{\mathcal K}^\dagger_{[t]}| + t = K$ naturally hold according to the definitions of $\mathcal{K}_{[t]}$, $\widetilde{\mathcal K}^\dagger_{[t]}$ and $\mathcal{K}^*_{[t]}$. Hence, we now have
        \begin{align}
          \mathsf{\Gamma}( \mathcal{K}^*_{[t]}, \mathcal{K}^\dagger_{[t]} ) =& \mathsf{\Gamma} ( \mathcal{K}_{[t]}, \mathcal{K}^*_{[t]}) + \mathsf{\Gamma} ( \widetilde{\mathcal K}^\dagger_{[t]}, \mathcal{K}^*_{[t]})  \geq \frac{t|\mathcal{K}^*_{[t]}|}{K(K-1)} \mathsf{\Gamma}(\mathcal{K}^*)
        \end{align}
        Considering $\mathsf{\Phi}'(\cdot)$ is a monotone submodular function, we have 
        \begin{align}
          \mathsf{\Phi}'\left(\mathcal{K}^*_{[t]} \bigcup \mathcal{K}^\dagger_{[t]}\right) - \mathsf{\Phi}'\left(\mathcal{K}^\dagger_{[t]}\right) \geq \mathsf{\Phi'}\left(\mathcal{K}^*\right) - \mathsf{\Phi}'\left(\mathcal{K}^\dagger\right)
        \end{align}
        since $\mathcal K^* \subseteq \mathcal{K}^*_{[t]} \bigcup \mathcal{K}_{[t]}$ and $\mathcal{K}^\dagger_{[t]} \subseteq \mathcal{K}^\dagger$. Moreover, according to \textbf{Lemma}~\ref{le:submodularpro}, we also have
        \begin{align}
          \sum_{i \in {\mathcal{K}}^*_{[t]}} \Delta\mathsf{\Phi}'_{i} \left( \mathcal{K}^\dagger_{[t]} \right) =&  \sum_{i \in {\mathcal{K}}^*_{[t]}} \left( \mathsf{\Phi}' \left( \{i\} \bigcup \mathcal{K}^\dagger_{[t]} \right) - \mathsf{\Phi}' \left( \mathcal{K}^\dagger_{[t]} \right) \right)  \nonumber\\
          \geq& \mathsf{\Phi}' \left( {\mathcal{K}}^*_{[t]} \bigcup \mathcal{K}^\dagger_{[t]} \right) - \mathsf{\Phi}' \left( \mathcal{K}^\dagger_{[t]} \right)  \nonumber\\
          \geq& \mathsf{\Phi'} \left( \mathcal{K}^* \right) - \mathsf{\Phi}' \left( \mathcal{K}^\dagger \right)
        \end{align}
        Then, we obtain
        \begin{align}
          &\sum_{i \in \mathcal{K}^*_{[t]}} \Delta\mathsf{F}'_{i} \left( \mathcal{K}^\dagger_{[t]} \right)  \nonumber\\
          =& \sum_{i \in \mathcal{K}^*_{[t]}} \left( \mathsf{F}' \left( \{i\} \bigcup \mathcal{K}^\dagger_{[t]} \right) - \mathsf{F}' \left( \mathcal{K}^\dagger_{[t]} \right) \right)  \nonumber\\
          =& \frac{1}{2} \sum_{i \in \mathcal{K}^*_{[t]}} \Delta\mathsf{\Phi}'_{i} \left( \mathcal{K}^\dagger_{[t]} \right) + \lambda \xi \mathsf{\Gamma} \left( \mathcal{K}^*_{[t]}, \mathcal{K}^\dagger_{[t]} \right)  \nonumber\\
          \geq& \frac{1}{2} \left( \mathsf{\Phi'} \left( \mathcal{K}^* \right) - \mathsf{\Phi}' \left( \mathcal{K}^\dagger \right) \right) + \lambda\xi \frac{t|\mathcal{K}^*_{[t]}|}{K(K-1)}\mathsf{\Gamma} \left( \mathcal{K}^* \right)
        \end{align}
        Let $i_{[t]}$ denote the service selected by our greedy algorithm in the $(t+1)$-th iteration such that $\Delta\mathsf{F}'_{i_{[t]}}(\mathcal{K}^\dagger_{[t]}) \geq \Delta\mathsf{F}'_{i}(\mathcal{K}^\dagger_{[t]})$ for any $i \in \mathcal{K}^*_{[t]}$. Therefore, we have
        \begin{align}
          &\Delta\mathsf{F}'_{i_{[t+1]}}(\mathcal{K}^\dagger_{[t]})  \nonumber\\
          \geq& \frac{1}{|\mathcal{K}^*_{[t]}|} \sum_{i \in \mathcal{K}^*_{[t]}} \Delta\mathsf{F}'_{i}(\mathcal{K}^\dagger_{[t]})  \nonumber\\
          \geq& \frac{1}{2|\mathcal{K}^*_{[t]}|} \left( \mathsf{\Phi'}(\mathcal K^*) - \mathsf{\Phi}'(\mathcal{K}^\dagger ) \right) + \lambda\xi \frac{t}{K(K-1)}\mathsf{\Gamma}(\mathcal K^*)  \nonumber\\
          \geq& \frac{1}{2K} \left( \mathsf{\Phi'}(\mathcal K^*) - \mathsf{\Phi}'(\mathcal{K}^\dagger ) \right) + \lambda\xi \frac{t}{2K(K-1)}\mathsf{\Gamma}(\mathcal K^*)
        \end{align}
        according to which, we can re-write $\mathsf{F}'(\mathcal{K}\dagger)$
        \begin{align}
          \mathsf{F}'(\mathcal{K}^\dagger) =& \sum^{K-1}_{t=0} \Delta\mathsf{F}'_{i_{[t+1]}}(\mathcal{K}^\dagger_{[t]})  \nonumber\\
          \geq& \frac{1}{2} \left( \mathsf{\Phi'}(\mathcal K^*) - \mathsf{\Phi}'(\mathcal{K}^\dagger ) \right) +  \frac{\lambda\xi}{2}\mathsf{\Gamma}(\mathcal K^*)
        \end{align}
        Since $\mathsf{F}'(\mathcal{K}^\dagger) = \frac{1}{2}\mathsf{\Phi}'(\mathcal{K}^\dagger) + \lambda\xi \mathsf{\Gamma}(\mathcal{K}^\dagger)$, we obtain
        \begin{align}
          &\mathsf{\Phi}'(\mathcal{K}^\dagger) + \lambda\xi \mathsf{\Gamma}(\mathcal{K}^\dagger) \geq \frac{1}{2} \mathsf{\Phi'}(\mathcal K^*) +  \frac{\lambda\xi}{2}\mathsf{\Gamma}(\mathcal K^*) = \frac{1}{2}\mathsf{F}(\mathcal{K}^*)
        \end{align}
        Finally,
        \begin{align}
          \mathsf{F}(\mathcal{K}^\dagger) =& \mathsf{\Phi}'(\mathcal{K}^\dagger) + \lambda\xi \mathsf{\Gamma}(\mathcal{K}^\dagger) \geq \frac{1}{2}\mathsf{F}(\mathcal{K}^*)
        \end{align}
      \end{proof}

      We have proved the approximation ratio of \textbf{Algorithm}~\ref{alg:greedy}, and we then analyze its time complexity in \textbf{Theorem}~\ref{thm:timecomplexity}. 
      \begin{theorem} \label{thm:timecomplexity}
        The time complexity of our top-$K$ service recommendation algorithm (see \textbf{Algorithm}~\ref{alg:greedy}) is $\mathcal{O}(M^2 K^3 + M N K^4)$.
      \end{theorem}
      \begin{proof}

        As demonstrated in \textbf{Algorithm}~\ref{alg:greedy}, in each iteration, it seeks for a service $i^\dagger \in \overline{\mathcal M}_{j^\dagger}$ such that $\mathsf{F}' \left( \mathcal{K} \bigcup \{i^\dagger\} \right)$ is maximized by $\mathcal{K} \bigcup \{i^\dagger\}$, where $\mathcal{K}$ is the set of services which are already selected. The above procedure is repeated until $K$ services is selected in total.

        Let $\mathcal{K}_{[t]} \subseteq \overline{\mathcal M}_{j^\dagger}$ denote the set of selected services in the $t$-th iteration and $K_{[t]} = |\mathcal{K}_{[t]} | = t$. In the $(t+1)$-th iteration, the time complexity of calculating $\mathsf{F}' \left( \mathcal{K}_{[t]} \bigcup \{i\} \right)$ for $i \in \overline{\mathcal M}_{j^\dagger} / \mathcal{K}_{[t]}$ relies on the ones of computing $\mathsf{Acc} \left( \mathcal{K}_{[t]} \bigcup \{i\} \right)$, $\alpha \left( \mathcal{K}_{[t]} \bigcup \{i\} \right)$ and $\beta \left( \mathcal{K}_{[t]} \bigcup \{i\} \right)$, as illustrated by (\ref{eq:fprime}). According to the definition of $\mathsf{Acc}$ shown in (\ref{eq:accuracy}), the time complexity of calculating $\mathsf{Acc}$ is $\mathcal{O} \left( K_{[t]} \right)$. Additionally, according to our definitions of $\alpha$ and $\beta$ as shown in (\ref{eq:expratio}) and (\ref{eq:dirdiv}), respectively, the time complexities are $\mathcal{O} \left( M K_{[t]} \right)$ and $\mathcal{O}\left( N K^2_{[t]} \right)$, respectively. Therefore, the time complexity of calculating $\mathsf{F}' \left( \mathcal{K}_{[t]} \bigcup \{i\} \right)$ in the the $(t+1)$-th iteration is $\mathcal{O}\left( M K_{[t]} + N K^2_{[t]} \right)$ and the one of calculating $i^\dagger = \arg\max_{i \in i \in \overline{\mathcal M}_{j^\dagger} / \mathcal{K}_{[t]}} \mathsf{F}'(\mathcal{K}_{[t]} \bigcup \{i\})$ is thus $\mathcal{O}\left( \left(M K_{[t]} + N K^2_{[t]} \right) \left( \overline{M}_{j^\dagger} - t \right) \right)$. Since $\mathcal{K}_{[t+1]} = \mathcal{K}_{[t]} \bigcup \{i^\dagger\}$, the total time complexity across the $K$ iterations is $\mathcal{O}(M^2 K^3 + M N K^4)$.
      \end{proof}

      Note that $K$ is usually considered as a constant for any target user, which is much smaller than $M$ and $N$. The actual time complexity of our algorithm can be re-written as $\mathcal{O} (M^2 + MN)$.

\section{Experiment}\label{sec:experiment}
  In this section, we first present our experiment settings and the metrics for evaluation in Sec.~\ref{ssec:setting}. We then introduce the state-of-the-art methods in \ref{ssec:refer} and compare them with our PDSR method in Sec.~\ref{ssec:results}. 

  \subsection{Experiment Settings and Metrics}  \label{ssec:setting}

    In this paper, we conduct our experiments on both the WS-DREAM dataset~\cite{wsdream} and the MovieLens dataset~\cite{mlens} to evaluate our PDSR method. The WS-DREAM dataset contains real-world QoS data, e.g., response time values, collected from $339$ users interacting with $5,825$ web services. To facilitate our comparison, each QoS value is normalized into a range of $[0,1]$. The MovieLens dataset is another commonly used dataset for studying service recommendation. It involves the ratings given by $6,040$ users for $3,952$ movies. We thus treat these ratings as the QoS values.
    For each user $j$ in WS-DREAM dataset and the one in MovieLens dataset, we randomly choose $\mathcal S_j \subset \mathcal M_j$ such that $S = |\mathcal S_j| = 15$, where $\mathcal{M}_j$ denotes the set of the services invoked by user $j$, and let $\mathcal S_j$ be the test dataset. Specifically, since MovieLens dataset is rather sparse, we take into account only the users who have no less than $25$ service usage records such that we have sufficient historical QoS data to make recommendation decisions for each target user by our method. We assume there are two platforms and divide the WS-DREAM dataset into two subsets such that the platforms have $135$ users and $204$ users, respectively. Likewise, the MovieLens dataset is divided such that the two platforms have $2,249$ users and $3,375$ users, respectively. We randomly pick up $15$ users on each platform as target users, to each of which, we deliver a recommendation list containing $K= 5$ services.

    We use the following four metrics to evaluate our PDSR method as well as the reference ones.
    \begin{itemize}
      \item \textit{Mean Absolute Error} (MAE) and \textit{Root Mean Squared Error} (RMSE). Since service recommendation highly rely on the predictions of the QoS values, we hereby adopt MAE and RMSE to measure the difference between our predictions and the actual values. Specifically, given a set of target users $\mathcal N^\dagger_r \subseteq \mathcal N$ on platform $r$, the MAE can be defined by
      \begin{align} \label{eq:mae}
        \mathsf{MAE} = \frac{1}{N^\dagger_r S} \sum_{j \in \mathcal N^\dagger_r} \sum_{i \in \mathcal S_j}  \left| d_{i,j,r} - d^*_{i,j,r} \right|
      \end{align}
      where $N^\dagger_r = | \mathcal N^\dagger_r |$ denotes the number of the target users on platform $r$, while $d_{i,j,r}$ and $d^*_{i,j,r}$ denote the predicted QoS value of service $i$ for user $j$ on platform $r$ and its ground truth, respectively. Similarly, the RMSE on platform $r$ can be calculated as
      \begin{align} \label{eq:rmse}
        \mathsf{RMSE} = \sqrt{\frac{1}{N^\dagger_r S} \sum_{j \in \mathcal N^\dagger_r} \sum_{i \in \mathcal S_j} \left( d_{i,j,r} - d^*_{i,j,r} \right)^2}
      \end{align}

      According to the above definitions, smaller MAE and RMSE indicate better prediction accuracy.
      \item \textit{Average Quality of Service} (AQoS). We employ AQoS to reflect users' actual interests in the recommended services. In particular, we define AQoS as
      \begin{equation} \label{eq:tq}
        \mathsf{AQoS} = \frac{1}{K N^\dagger_r} \sum_{j \in \mathcal N^\dagger_r} \sum_{i \in \mathcal{K}^\dagger_j} d^*_{i,j,r}
      \end{equation}
      where $\mathcal{K}^\dagger_j$ denotes the recommendation list for user $j \in \mathcal N^\dagger_r$ on platform $r$. Higher AQoS implies that the diverse services we recommend to target users have better actual quality.
      \item \textit{Inter-List Diversity} (ILD). ILD is a commonly used metric to measure the diversity of a recommendation list \cite{ZhangWHCDY-ICWS19}. Specifically,
      \begin{align}
        \mathsf{ILD} = \frac{1}{N^\dagger_r} \sum_{j \in \mathcal N^\dagger_r} \left( \frac{ \sum_{i,i'\in \mathcal{K}^\dagger_j, i \neq i'}  \mathsf{J}^*_{i,i',r} }{K(K-1)} \right)
      \end{align}
      where $\mathsf{J}^*_{i,i',r} = \mathsf{J}\left({\vec{d}}^*_{i,r},\vec{d}^*_{i',r}\right)$, as defined in Eq.~\ref{eq:jaccard}, $\mathsf{J}\left({\vec{d}}^*_{i,r},\vec{d}^*_{i',r}\right)$ represents the \textit{Jaccard dissimilarity} between services $i, i'$. 
      {The higher the $\mathsf{ILD}$ is, the higher the recommendation diversity is.}
    \end{itemize}

  \subsection{Reference Methods}  \label{ssec:refer}
    In this section, we compare our method with the following six state-of-the-art ones.
    \begin{itemize}
      \item \textbf{BPR method}~\cite{RendleFGT-UAI09}: The \textit{Bayesian Personalized Ranking} (BPR) method applies a Bayesian analysis to the personalized ranking problem. In particular, a maximum posterior estimator is derived to predict a personalized ranking.
      \item \textbf{MPR method}~\cite{YuZYWWLC-CIKM18}: The \textit{Multiple Pairwise Ranking} (MPR) method deeply exploits the unobserved item feedback to perform multiple pairwise ranking. Specifically, it divides the unobserved items into different parts to exploit the preference difference among multiple pairs of items, which is thus thought of as a multiple pairwise model.
      \item \textbf{EF method}~\cite{LathaN-Physica19}: In the \textit{Entropy Fusion} (EF) method based on CF, accuracy-diversity trade-off are leveraged by taking into account both user exposure diversity and item concordance.
      \item \textbf{MF-based method}~\cite{WangWLWLQ-WWW22}: The \textit{Matrix Factoring} (MF)-based method considers both the explicit and implicit information during the matrix factoring process;popularity bias with a weighting mechanism and neighborhood information are used to diversifying the recommendation results.

      \item \textbf{DPP-based method} ~\cite{ChenZZ-NIPS18} : Since DPP is a powerful tool for modeling diversity, \cite{ChenZZ-NIPS18} proposes an algorithm to accelerate the greedy MAP inference for DPP, and the algorithm can be applied to the task of diversified recommendation.

      \item \textbf{LSH-based method}~\cite{WangZWYKQ-KBS20}: In this method, the LSH mechanism is used to calculate both similarity and diversity. Additionally, thanks to the LSH mechanism, this method can be adapted to utilize across-platform data with privacy preserved. 
    \end{itemize}
    Note that the BPR method and the MPR method only aim at improving recommendation accuracy, while the other four methods investigate the accuracy-diversity trade-off. Moreover, since the first five of the above methods do not take into account the privacy issue, we assume that they utilize only the original data on single platforms.

  \subsection{Experiment Results}  \label{ssec:results}
    \subsubsection{Comparison with Different Methods} \label{sssec:comparison}

      We first compare our PDSR method with the six reference ones under the metrics including MAE, RMSE, AQoS, and ILD. We let $H_1 = H_2 = 3 $, $T= 9$, $\lambda= 0.1$ and $\xi= 0.3 $ for Platform 1 and $H_1 = H_2 = 3 $, $T= 9$ and $\lambda = \xi= 0.3$ for Platform 2, when performing our experiments on the WS-DREAM dataset. As for the MovieLens dataset, we let $H_1 = H_2 = 4 $, $T= 9$, $\lambda= 0.1$ and $\xi= 0.3 $ for Platform 1 and $H_1 = H_2 = 5 $, $T= 9$, $\lambda= 0.1$ and $\xi= 0.2$ for Platform 2. For each experiment data we report, we repeat the experiments fifty times and take an average over the experiment results.
      \begin{table*}
      \caption{Comparison results in terms of MAE, RMSE, AQoS and ILD.}
      \label{tb:comparison}
      \centering
      \setlength{\tabcolsep}{3.5mm}
      \renewcommand{\arraystretch}{1.4}
      \begin{tabular}{c|c|c|c|c|c|c|c|c|c} 
        \hline
        \multirow{2}{*}{Dataset}  & \multirow{2}{*}{Methods} 
        & \multicolumn{4}{c|}{Platform 1} & \multicolumn{4}{c}{Platform 2}  \\
        \cline{3-10}
        &  & MAE    & RMSE     & AQoS   & ILD  
           & MAE   & RMSE    & AQoS   & ILD     \\
        \hline
        \multirow{7}{*}{WS-DREAM}  
        & BPR  & 0.9354  & 1.0565  & 0.8945  & 0.4454 
               & 0.9308  & 1.0028  & 0.9123  & 0.4454     \\
        & MRP  & 1.486  & 1.8189  & 0.8952  & 0.4455  
               & 1.1678  & 1.3974  & 0.9142  & 0.4453     \\
        & EF  & 0.0885   &  0.1301  & 0.9025  & 0.4454  
              & 0.0748  & 0.1097  & 0.9186  & 0.4454       \\
        & MF-based  & 0.509   & 0.5235   & 0.9035  & 0.442             & 0.4893   & 0.506 & 0.9184   & 0.4435          \\
        & DPP    & 0.0783  & 0.1158  & 0.9025  & 0.4454  
                & 0.0703  & 0.1038   & 0.9089  & 0.4453      \\
        & LSH-based  & 0.0782  & 0.1158  & 0.8989  & 0.4471 
                    & 0.0703  & 0.1039  & 0.8955  & 0.4477   \\
        & PDSR  & {0.0782}  & {0.1158}  &{0.9042}  & {0.462}          
                & {0.0703}  & {0.1038}  &{0.9191}  & {0.4603}   \\   
        \hline
        \multirow{7}{*}{MovieLens} 
        & BPR  & 1.7207  & 2.1401  & {3.944}  & 0.3843  
               & 1.8111  & 2.2051  & {3.9676}  & 0.3818    \\
        & MPR  & 1.7553  & 2.2217  & 3.8969  & 0.3872  
               & 1.7543  & 2.2133  & 3.9298  & 0.3851   \\
        & EF  & 1.1173  & 1.311   & 3.6382  & 0.4165  
              & 1.2378  & 1.3332  & 3.75    & 0.4263    \\
        & MF-based  & 1.9795  & 2.1632  & 3.938  & 0.3883  & 2.1496  & 2.3298  & 3.9188  & 0.3971         \\
        & DPP   & 0.8748   & 1.044  & 3.856  & 0.4256     
                & 0.8713   & 1.0379   & 3.8293  & 0.4215      \\
        & LSH-based  & 0.8638  & 1.0565  & 3.8021  & 0.4369 
                     & 0.8405  & 1.0422  & 3.7989  & 0.4279    \\
        & PDSR  & {0.8443}  & {1.0248}  & {3.9}  & {0.4324}  
                & {0.8326}  & {1.0189}  &{3.9138}  & {0.4274}    \\
        
      \hline
      \end{tabular}
      \end{table*}

      As shown in Table~\ref{tb:comparison}, on the WS-DREAM dataset, our PDSR method has smaller MAE and RMSE than the reference ones (especially, the BPR, MRP, EF, and MF-based methods) on both the two platforms. This implies that our PDSR method makes QoS predictions more accurately, thanks to the privacy-preserved data sharing across the different platforms. Furthermore, compared with the BRP and MPR methods, our PDSR method takes into account recommendation diversity, and can explore potential services; therefore, it has higher AQoS and higher ILD. Additionally, the AQoS and ILD of our PDSR method are higher than the EF, MF-based and DPP methods where the recommendation diversity is also considered, since the PDSR method not only predicts the missed QoS values more accurately but also enables higher recommendation diversity through data sharing. Although the LSH-based method is extended to enable the data sharing with privacy preserved, it adopts the LSH mechanism directly to leverage the accuracy-diversity trade-off, while our method elaborately defines the accuracy-diversity metric upon the similarity graph as shown in Sec.~\ref{ssec:recomm}. Hence, our method has higher AQoS and ILD than the LSH-based one.

      The advantage of our privacy-preserved data sharing in prediction accuracy also can be revealed through the experiment results on the MovieLens dataset. Specifically, the MAE and RMSE of our PDSR method are smaller than the other reference ones on both of the platforms. Another lesson we learn from the experiment results is that, excessively high diversity (i.e., ILD) may result in decreased AQoS. For example, our PDSR method has higher recommendation diversity than the MF-based methods on Platform 1, but the AQoS of our method is slightly smaller than the one of the MF-based method. We will discuss about the trade-off between AQoS and ILD later in Sec.~\ref{sssec:lambda}.

      We also report the experiment results in terms of time cost. Since the reference methods (e.g., the first five ones) do not consider the data sharing across the two platforms, we omit the communication overhead induced by the data sharing for the purpose of comparability. 
      We conduct our experiments on a desktop equipped with 2.40 GHz CPU and 16 GB RAM. As illustrated in Table~\ref{tb:time cost}, our PDSR method has much smaller time cost than the first five reference methods on both the two datasets. As for the LSH-based method, since it directly adopts LSH to measure both similarity and diversity, it results in slightly less time cost. Nevertheless, considering the advantage of our method under another metrics (e.g., MAE, RMSE, AQoS, and ILD as shown in Table~\ref{tb:comparison}), this is the price we have to pay.
      \begin{table*}
      \caption{Time cost.}
      \label{tb:time cost}
      \centering
      \setlength{\tabcolsep}{0.4mm}
      \renewcommand{\arraystretch}{1.5}
      \begin{tabular}{c|c|c|c|c|c|c|c|c|c|c|c|c|c|c} 
        \hline
        \multirow{2}{*}{Dataset}   
        & \multicolumn{7}{c|}{Platform 1} &\multicolumn{7}{c}{Platform 2}  \\
        \cline{2-15}
          & BPR   & MPR  & EF  & MF-based  & DPP  & LSH-based   & PDSR   
          & BPR   & MPR  & EF  & MF-based  & DPP  & LSH-based   & PDSR   \\
        \hline
        \multirow{1}{*}{WS-DREAM}  
          & 18.482  & 78.919  & 176.731  & 79.658
          & 89.760  & 11.210  & {16.803}    
          & 18.524  & 143.78  4& 185.252   & 98.274
          & 76.344  & 11.268   & {16.825}\\
        \hline
        \multirow{1}{*}{MovieLens} 
          & 84.061  & 308.796  & 183.592  & 637.354  
          & 18.244  & 6.366   & {10.773}  
          & 126.608   & 599.978  & 344.063  & 1007.899
          & 25.178    & 6.2857   & {10.963}\\
        \hline
      \end{tabular}
      \end{table*}

    \subsubsection{Fine Tuning Locality-Sensitive Hashing}  \label{sssec:handt}

\begin{figure}[htb!]
      \begin{center}
        \parbox{.9\columnwidth}{\center\includegraphics[width=.75\columnwidth]{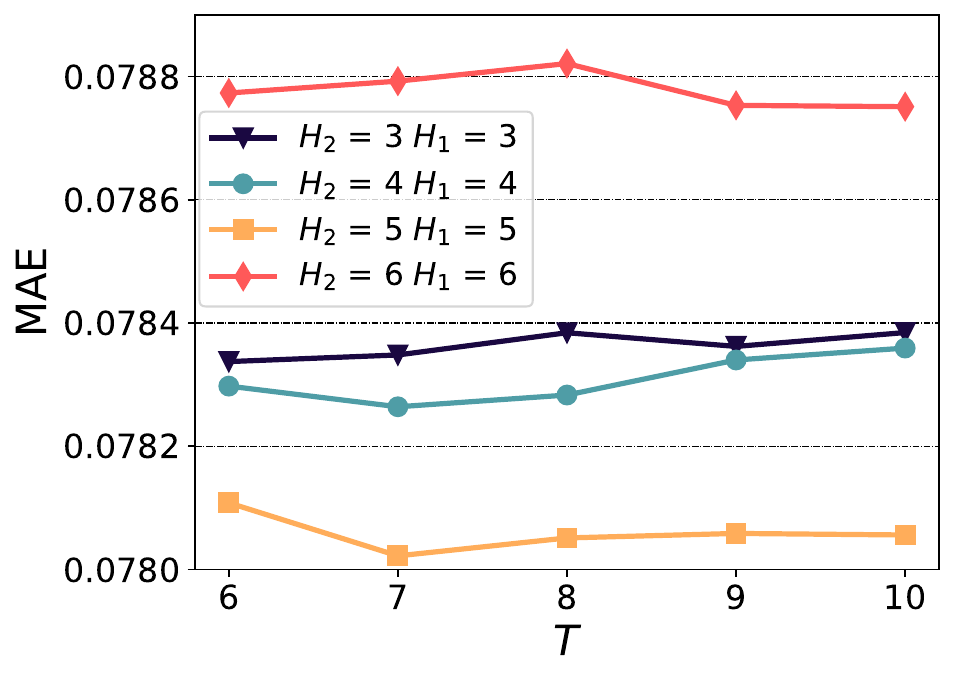}}
        \parbox{.9\columnwidth}{\center\scriptsize(a) MAE} 
        \parbox{.9\columnwidth}{\center\includegraphics[width=.75\columnwidth]{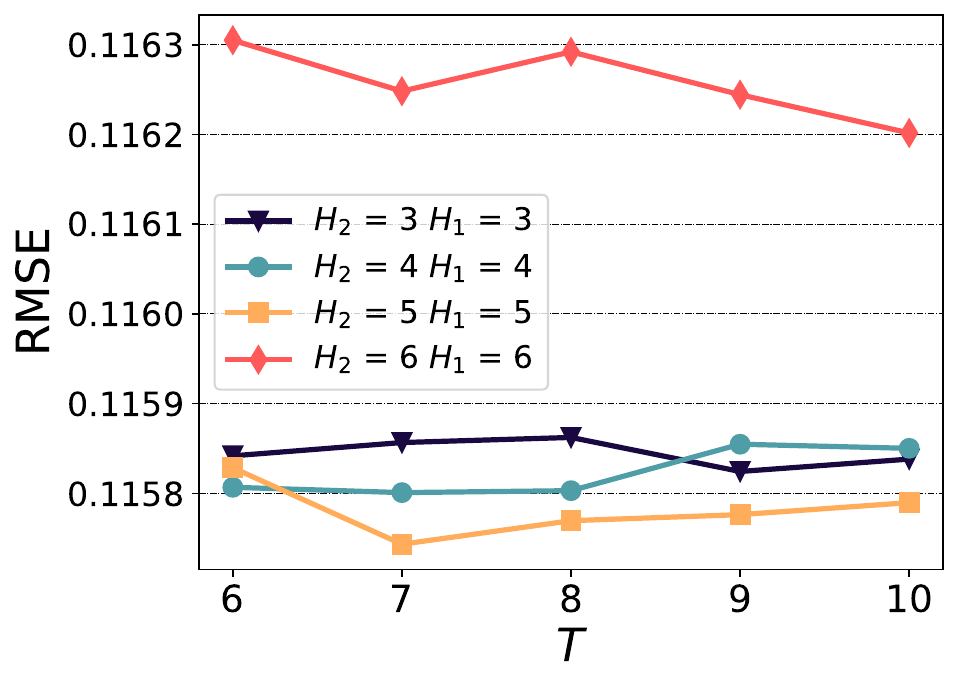}}
        \parbox{.9\columnwidth}{\center\scriptsize(b) RMSE}
        \caption{Prediction accuracy with different values of $H$ and $T$.}
        \label{fig:finetuning}
      \end{center}
      \end{figure}

     Recall that, in our method, we index the services using $H_r$ hash functions on each platform $r$ (see \textbf{Algorithm}~\ref{alg:servindex}) and measure the similarity among the services by calling the subroutine of service indexing $T$ times across the different platforms (see \textbf{Algorithm}~\ref{alg:graph}). We hereby reveal how $H_r$ (where $r=1,2$ in our case) impacts the performance of our method in terms of QoS prediction. Specifically, let $H_r \in \{ 3, 4, 5, 6 \}$ for $r = 1,2$ and $T \in \{ 6, 7, 8, 9, 10\}$. The results are shown in Fig.~\ref{fig:finetuning}. Since we get similar results for the two datasets on both platforms, we report only the ones obtained with the WS-DREAM dataset on Platform $1$.

     According to \textbf{Algorithm}~\ref{alg:servindex}, the number of the hash functions and the number of the hash tables determine the ``resolution'' at which we characterize the similarity. It is shown in Fig.~\ref{fig:finetuning} that, fixing the number of the hash tables, i.e., $T$, when the number of hash functions used in \textbf{Algorithm}~\ref{alg:servindex}, i.e., $H_r$, is increased, the MAE and RMSE of our method is first decreased but then increased. When $H_r$ is smaller, introducing more hash functions helps us to characterize the service similarity more accurately. However, we may underestimate the service similarity when there are too many hash functions. Specifically, it is more likely for the services to have different hash values in this case, even they are similar to each other. Furthermore, $T$ actually has similar impact on the MAE and RMSE, as shown in \textbf{Theorem}~\ref{thm:eff}. By and large, the actual impact of $T$ is slighter than the one of $H_r$, especially when $H_r$ is set such that a larger prediction error is induced (e.g., when $H_r = 3, 4, 6$). Nevertheless, when $H_r=5$, we are still able to subtly reduce the MAE and RMSE of our prediction by increasing $T$. However, recklessly increasing $T$ finally results in a growth of the prediction error.  

    \subsubsection{An Accuracy-Diversity Trade-off}  \label{sssec:lambda}

\begin{figure}[htb!]
      \begin{center}
        \parbox{.9\columnwidth}{\center\includegraphics[width=.75\columnwidth]{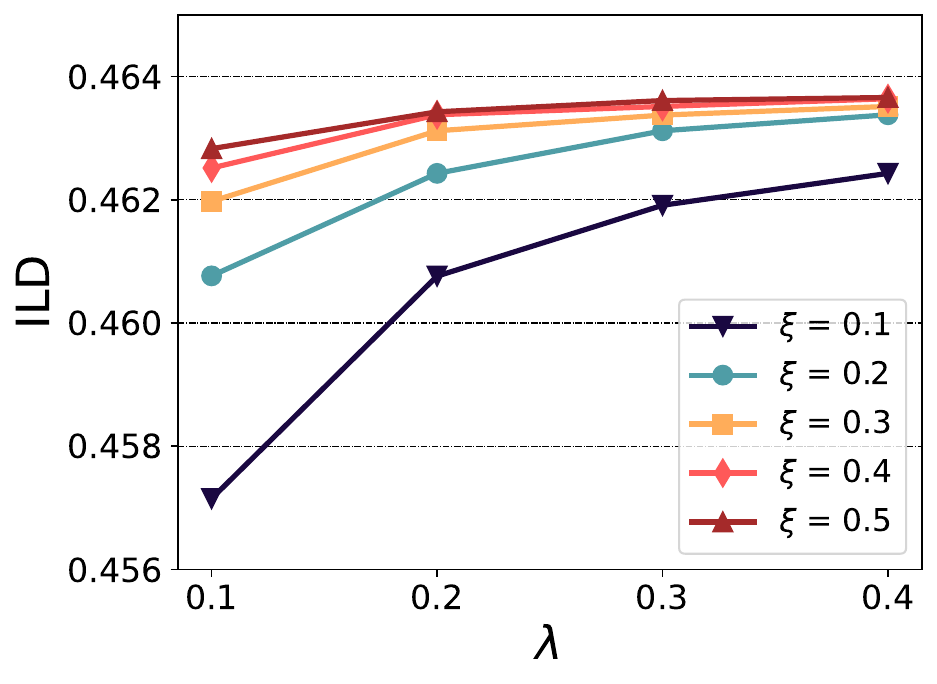}}
        \parbox{.9\columnwidth}{\center\scriptsize(a) ILD}
        \parbox{.9\columnwidth}{\center\includegraphics[width=.75\columnwidth]{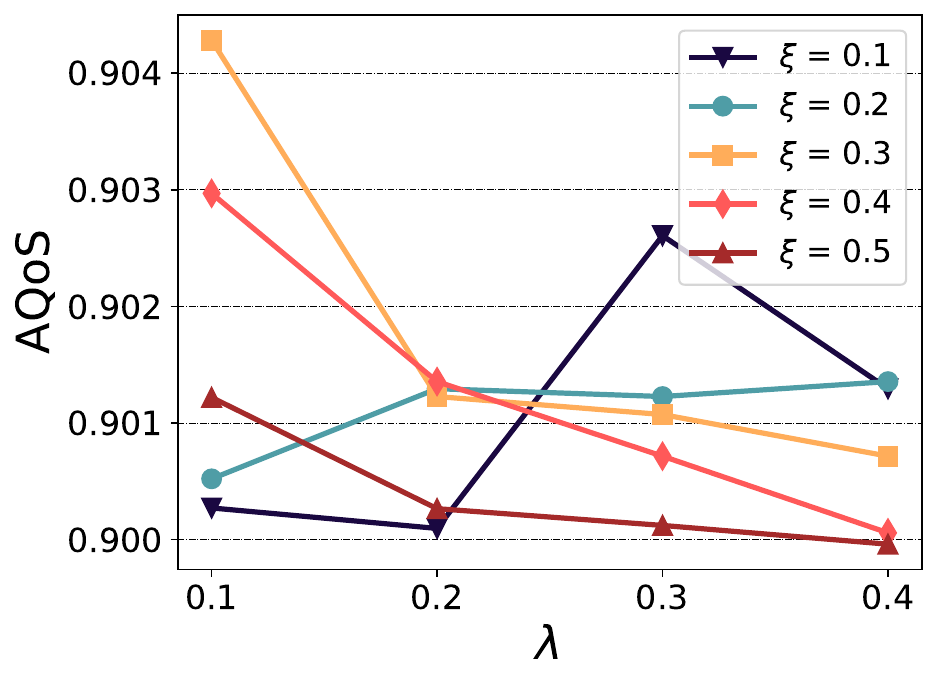}}
        \parbox{.9\columnwidth}{\center\scriptsize(b) AQoS}
        \caption{Trade-off between accuracy and diversity.}
        \label{fig:accdiv}
      \end{center}
      \end{figure}

      In our PDSR method, we leverage $\lambda$ to make a trade-off between accuracy and diversity. We hereby report only the AQoS and ILD values obtained on Platform 1 using the WS-DREAM dataset in Fig.~\ref{fig:accdiv}, since the results with the MovieLens dataset and the ones on Platform 2 are rather similar. We let $\lambda = 0.1, 0.2, 0.3, 0.4$ and vary $\xi = 0.1, 0.2, 0.3, 0.4, 0.5$ to illustrate how the composition of the diversity impact the accuracy-diversity trade-off. It is demonstrated in Fig.~\ref{fig:accdiv} (a) that we obtain higher ILD by increasing $\lambda$. As mentioned in Sec.~\ref{sec:intro}, taking into account diversity help us to explore the potential services with higher AQoS. Therefore, as shown in Fig.~\ref{fig:accdiv} (b), when $\xi=0.2$, we obtain higher AQoS by increasing $\lambda$ from $0.1$ to $0.2$. Nevertheless, pursuing the recommendation diversity excessively may impair the quality of our service recommendation. For example, when $\xi=0.3, 0.4, 0.5$, increasing $\lambda$ leads to recommendation results with lower AQoS.

\section{Conclusion}  \label{sec:conclusion}
  In this paper, we have proposed a \textit{Privacy-preserving Diversified Service Recommendation} (PDSR) method. The method leverages LSH mechanism to enable data sharing among different data sources (or platforms) with privacy preserved. Specifically, according to the LSH values, we can construct a graph to characterize the similarity among the services, based on which we predict missing QoS values accurately. Furthermore, we have designed a novel diversity metric and proposed a $2$-approximation algorithm to select $K$ services by maximizing the objective of accuracy-diversity trade-off. We have performed extensive experiments on real datasets to verify the efficacy of our PDSR method.

  Whereas our PDSR method currently exploit diverse services with diverse qualities to fulfill users' different demands, we are on the way of extending our definition of diversity by taking into account more attribute of the services. For example, we can consider the functional features of the services such that the services with different functional features have higher diversity measurements. Moreover, the users may be interested in the services which provide a combination of specific functions. In this case, to fulfill the users' requirement on service functions, we need to recommend multiple sets of services such that each of the sets can provide the specific functions but has diversified QoS measurement.

\bibliographystyle{unsrt}
\bibliography{pdsr.bib}  

\end{document}